\documentclass[11pt]{article}
\usepackage[T1]{fontenc}
\usepackage[dvipsnames]{xcolor}
\usepackage{graphicx, subfigure}
\usepackage[top=1in,bottom=1in,left=1in,right=1in]{geometry}
\usepackage[sort&compress,round]{natbib} 
\usepackage{amsfonts, amsmath, amssymb, amsthm, constants, bbm}
\usepackage{MnSymbol}
\usepackage{mathtools}
\usepackage{enumitem}
\usepackage{caption}
\usepackage{algpseudocode}
\usepackage{pifont}   % for \cmark

\usepackage{algorithm}
\usepackage{booktabs}       % \toprule, \midrule, \bottomrule
\usepackage{tikz}
\usetikzlibrary{positioning,arrows.meta}
\usepackage{cellspace}
\usepackage{natbib}
\usepackage{makecell}
\setcellgapes{4pt}

\usepackage{tabularray}
\UseTblrLibrary{booktabs}

\definecolor{darkred}{RGB}{100,0,0}
\definecolor{darkgreen}{RGB}{0,100,0}
\definecolor{darkblue}{RGB}{0,0,150}
\definecolor{citecol}{RGB}{30,80,150}
\definecolor{tabcol}{RGB}{200,230,255}
\definecolor{coolblack}{RGB}{40, 40, 45}
\definecolor{coolwhite}{RGB}{160, 160, 160}
\definecolor{coolblue}{RGB}{65, 130, 205}
\definecolor{coolgreen}{RGB}{8, 138, 94}
\definecolor{study1}{HTML}{E63946} % Crimson red
\definecolor{study2}{HTML}{2196F3} % Ocean blue  
\definecolor{study3}{HTML}{FF9800} % Amber orange
\usepackage{colortbl}

\usepackage{hyperref}
\hypersetup{colorlinks=true, linkcolor=coolgreen, citecolor=coolgreen, urlcolor=darkblue}
\usepackage{url}

\newtheorem{prp}{Proposition}

\newtheorem{cor}{Corollary}

\newtheorem{ass}{Assumption}

\theoremstyle{remark}
\newtheorem{rem}{Remark}

\def\beq{\begin{equation}} % \setcounter{equation}{1}}
\def\eeq{\end{equation}}
\def\beqn{\begin{eqnarray*}}
\def\eeqn{\end{eqnarray*}}
\def\Bal{\begin{align}}
\def\Eal{\end{align}}
\def\Bitem{\begin{itemize}\setlength{\itemsep}{.2in}}
\def\bitem{\begin{itemize}\setlength{\itemsep}{.05in}}
\def\eitem{\end{itemize}}
\def\blatin{\begin{enumerate}\setlength{\itemsep}{.05in}\renewcommand{\labelenumi}{\roman{enumi}.}}
\def\elatin{\end{enumerate}}
\def\Benum{\begin{enumerate}\setlength{\itemsep}{.2in}}
\def\benum{\begin{enumerate}\setlength{\itemsep}{.05in}}
\def\eenum{\end{enumerate}}
\def\bmult{\begin{multline*}}
\def\emult{\end{multline*}}
\def\bcenter{\begin{center}}
\def\ecenter{\end{center}}
\def\bframe{\begin{frame}}
\def\eframe{\end{frame}}

\def\cA{\mathcal{A}}

\def\cE{\mathcal{E}}

\def\cK{\mathcal{K}}

\def\cN{\mathcal{N}}

\def\cP{\mathcal{P}}

\def\cX{\mathcal{X}}

\def\bP{\mathbf{P}}

\def\bd{\mathbf{d}}

\def\bn{\mathbf{n}}

\def\bp{\mathbf{p}}
\def\bq{\mathbf{q}}

\def\bbE{\mathbb{E}}

\def\bbP{\mathbb{P}}

\def\bbR{\mathbb{R}}

\def\ind{\mathbbm{1}}

\newcommand{\E}{\operatorname{\mathbb{E}}}

\newcommand{\Var}{\operatorname{Var}}

\newcommand{\indep}{\mathrel{\upmodels}}

\let\lac\{
\let\rac\}

\def\1{\mathbbm{1}}

\newcommand{\ve}{\varepsilon}

\newcommand*\diff{\mathop{}\!\mathrm{d}}
\newcommand\wt{\widetilde}

\usepackage[textsize=tiny]{todonotes}
\setuptodonotes{inline}

\title{Causal Perspectives on Network Meta-Analysis}
\author{Ahmed Boughdiri\footnotemark[2] \and Francisco Andrade\footnotemark[2] \and Clément Berenfeld\footnotemark[2] \and Julie Josse\footnotemark[2]}
\date{}

\begin{document}

\thispagestyle{empty}

\footnotetext[2]{PreMeDICaL, Inria, Inserm, Université de Montpellier, Montpellier, France. \\ 
Corresponding author: \href{mailto:julie.josse@inria.fr}{\tt julie.josse@inria.fr}}

\maketitle

\begin{abstract}

Pairwise and network meta-analyses occupy the highest tier of evidence-based medicine and routinely inform clinical guidelines and healthcare decision-making. Current approaches typically aggregate study-level treatment effects to obtain an overall estimate. We argue that the causal estimand should come first, with the aggregation derived only afterwards: the target population and the relevant sources of between-study heterogeneity should be explicitly defined before deriving the aggregation required for identification. This shift in perspective fundamentally changes both the estimands and the methodology.

We develop a unified causal framework for pairwise and network meta-analysis based on aggregate data. By defining treatment effects with respect to a clinically meaningful target population, for example, the average population represented by the contributing trials, and accounting for heterogeneity induced by treatment-effect modifiers and center effects, we show that identification naturally leads to arm-level aggregation. In the network setting, this causal formulation departs fundamentally from the conventional contrast-based paradigm: arm-level aggregation emerges from the causal formulation rather than from a modeling choice, and treatment effects are identified without relying on the treatment network itself. This perspective provides an additional conceptual argument in the long-standing contrast-based versus arm-based debate.

Numerical studies show that the proposed estimators target well-defined causal effects, whereas the causal interpretation of conventional approaches remains unclear. Although both approaches often produce similar estimates, we identify settings in which they diverge, with potentially important implications for the interpretation of meta-analytic evidence.

\end{abstract}

%\tableofcontents

\section{Introduction}

\paragraph{Motivations.} Meta-analyses occupy the highest level of the evidence hierarchy in evidence-based medicine and play a central role in informing decisions by health authorities and in particular  health technology assessment (HTA), particularly regarding drug reimbursement and market access. However, meta-analyses still face important methodological challenges, particularly when accounting for heterogeneity across trials. Moreover, they lack a causal inference perspective which implies that the target population for which the treatment effect is estimated is not explicit. %\medskip

Recently, \cite{berenfeld2025causal} introduced a causal framework for meta-analysis based on aggregated data that addresses several of these limitations. Their approach clarifies which forms of heterogeneity can be accommodated and, crucially, makes the \textbf{target population explicit}. This population may correspond, for example, to the average population across the included trials or to a weighted combination of trial populations. Such flexibility allows the analysis to reflect specific decision-making contexts—for instance, by assigning greater weight to a French trial when evaluating whether a treatment should be introduced into the French healthcare system.
The authors further characterize the conditions under which classical and causal meta-analyses coincide, as well as situations in which they may yield  different—sometimes even opposite—conclusions regarding treatment effects. Importantly, they demonstrate that, among commonly used effect measures, only the risk difference admits a causal interpretation within this framework. It means that there is not a well-defined target population in the other cases.  %\medskip

These new approaches can be framed within the estimand framework \citep{Kahan2024EstimandsPrimer}, while remaining fully compatible with the PICO (Population, Intervention, Comparator, Outcome) framework. In particular, the “P” becomes more explicitly defined, not only as the populations enrolled in each trial, but as the target population over which the treatment effect is to be estimated. %\medskip

From a methodological standpoint, the proposed causal approach is strikingly simple. In practice, it amounts to performing arm-level aggregation while applying weights that differ from those used in conventional meta-analytic methods. This framework opens new perspectives and raises important questions regarding the potential value of causal thinking in more complex meta-analytic settings, such as network meta-analysis, where the relative merits of contrast-based versus arm-based approaches have long been debated \citep{white2019armcontrast, hong2016bayesian, dias2016absolute}. %\medskip

Network meta-analysis \citep{chaimani2019network}  (NMA) synthesizes evidence from multiple randomized trials comparing different interventions for the same condition. Individual trials often evaluate only a subset of treatments, leaving some comparisons unobserved. NMA combines direct comparisons (within trials) and indirect comparisons (via a common comparator across trials) to estimate relative treatment effects even for interventions never compared head-to-head. It also enables treatment ranking using metrics such as SUCRA \citep{salanti2011rankings} or p-scores \citep{Rucker2015Ranking}. Key assumptions underpinning NMA include homogeneity of treatment effects, consistency between direct and indirect evidence, and transitivity across trials with comparable populations and effect modifiers. \medskip

\paragraph{Contributions and organization of the paper.} 

The primary objective of this paper is to develop a causal framework for NMA, in which the target population, causal estimands, and sources of heterogeneity are explicitly defined. To motivate this framework, we first review the classical approaches to pairwise and network meta-analysis in Section~\ref{sec:Reminders}. We then revisit the causal perspective on pairwise meta-analysis introduced by \cite{berenfeld2025causal} in Section~\ref{sec:causalpair}. Our first contribution is to extend their framework by accounting for an additional source of heterogeneity arising from center effects. Beyond differences in patient populations across studies, treatment effects may vary because of differences in clinical practices, operator expertise, equipment quality, or other center-specific characteristics. To capture this structure, we introduce a hierarchical modeling approach that jointly accounts for population and center-level heterogeneity.

Building on this causal pairwise framework, Section~\ref{sec:causalnma} develops a new causal approach to NMA. We adopt a progressive construction, starting from the idealized setting where all studies sample the same population and no center effect is present, before considering the more realistic setting of heterogeneous populations and center-specific effects. Our second contribution is to show that, under a causal formulation, the traditional treatment network graph is no longer a fundamental object for estimation. Consequently, the usual transitivity assumption is not used as a structural component of the methodology. This perspective leads to a novel and remarkably simple estimator for NMA, equipped with a clear causal interpretation in terms of a well-defined target population and explicit sources of heterogeneity. Finally, in Sections~\ref{sec:simulations} and \ref{sec:realworld}, we evaluate the proposed methods through simulation studies and real meta-analyses and compare their performance with existing approaches. 

\medskip
\paragraph{Related works.}

In recent years, there has been a growing interest in so-called causally interpretable meta-analysis (CIMA) \citep{dahabreh2020toward} for pairwise meta-analysis. However, in this literature, the term meta-analysis refers primarily to problems of generalizability and transportability, where individual participant data (IPD) from one or several randomized trials are used to estimate treatment effects in a prespecified external target population. 
More recently, \cite{shi2026causally} extended this framework to settings where IPD are unavailable and only  study-level summaries of patient covariates (e.g., baseline characteristics reported in Table~1) and subgroup-specific treatment effect estimates are available. These data are used to estimate a conditional treatment effect model and subsequently transport treatment effects to a target population characterized by known covariate characteristics.

Closest to our approach for pairwise meta-analysis, in terms of the data considered, is the work of \cite{zhang2026causal}, who propose a causal framework for aggregated data accounting for trial-level effects. In contrast, \cite{berenfeld2025causal} focus on the causal interpretation of standard meta-analytic estimators, showing that some lack a causal meaning and characterizing when classical and causal analyses diverge. Our hierarchical formulation of study-level effects is instead motivated by extensions to network meta-analysis.
 
For network meta-analysis, as far as we know, there is only the work of \cite{Schnitzer2016CausalNMA} who consider a causal framework where they define a target population of interest, referred to as a metapopulation, which encompasses the individual superpopulations underlying each study.
However, their primary focus is on adjusting for study-level confounding. Specifically, they assume the existence of trial-level features that influence both the selection of treatments evaluated in each study (including the choice of treatments and the number of arms) and the outcomes. To address this, they propose the use of doubly robust estimators and targeted maximum likelihood estimation (TMLE). 

Even though incorporating trial-level covariates, individual patient data, or subgroup information could refine the analysis, relax some assumptions, and allow transport of treatment effects to other populations, our work differs from previous approaches by deliberately considering a minimal setting based solely on the simplest aggregated data routinely available for both classical pairwise and network meta-analyses. Our aim is to clarify what can be learned from standard meta-analytic data alone, in terms of target populations, treatment effect heterogeneity, and the assumptions required for causal interpretation. This perspective reflects the practical reality that evidence synthesis is often conducted after trial completion and therefore relies on secondary data, typically with limited control over data collection and without access to individual-level information \citep{RemiroAzocar2025EstimandsMeta}.

\section{Reminders on classical (network) meta-analysis} \label{sec:Reminders}

\subsection{Pairwise meta-analysis}\label{sub_sec:pairwise_meta}

Pairwise meta-analysis combines results from studies comparing the effect of two interventions, say treatment $A=1$ versus treatment $A=0$, on the same outcome, that we denote by $Y$. We focus for the sake of simplicity on binary outcome $Y=0$ and $Y=1$. Each study $k \in [K]$ reports the values  $n^{ay}_k$ of the total number of individuals in this study taking treatment $A=a$ and with outcome $Y=y$, as in Table~\ref{tab:rct2}. We also let $n^{a}_k$ be the total number of individuals in study $k$ taking treatment $A=a$, and $n_k$ be the total number  of individuals in study $k$.

\begin{table}[h]
\centering
\begin{tabular}{c|cc}
         & $Y = 1$    & $Y = 0$    \\ \hline
$A = 1$  & $n_k^{11}$ & $n_k^{10}$ \\
$A = 0$  & $n_k^{01}$ & $n_k^{00}$\\
\end{tabular}
\caption{A typical table summarizing the finding of the RCT $k \in [K]$.}
\label{tab:rct2}
\end{table}

The synthesized effect is usually a relative contrast, such as (log) odds-ratio, (log) risk-ratios, or an absolute one like the risk difference between the two treatments. We will denote by $\hat \theta_k$ the estimated contrast in study $k$. This contrast is a function of the two absolute treatment effects $\hat\psi_k^a$ (with $a \in \{0,1\}$) reported in study $k$:
$$
\hat \theta_k = h(\hat\psi_k^1)-h(\hat\psi_k^0) \quad \text{where} \quad \hat\psi_k^a = \frac{n_k^{a1}}{n_k^a},
$$
and $h : [0,1] \to \bbR$ is a link function. 
Typical link functions include $h(x)=x$ (risk difference), $h(x)=\log(x)$ (log risk-ratio), or $h(x)=\log(x/(1-x))$ (log odds-ratio). The empirical contrast $\hat\theta_k$ always comes with a measure of uncertainty $\hat\sigma_k^2$ stemming from the normal approximation of the limiting law of $\hat\theta_k$ as $n_k$ goes to infinity. For instance, in the case of the log odds-ratio, we find that,
$$
\hat\theta_k := \log \left\{\frac{n^{11}_k}{n_k^1-n_k^{11}} \times \frac{n^0_k-n^{01}_k}{n_k^{01}}\right\} \quad \text{and} \quad \hat\sigma_k^2 = \frac{1}{n^{11}_k}-\frac{1}{n^{1}_k}+\frac{1}{n^{01}_k}-\frac{1}{n^{0}_k}.
$$

\paragraph{Fixed- vs random-effects models. }We let $\theta_k$ be the (true) contrast in study $k$, and $\sigma_k^2$ be the true within-study variance. Classical methods resort to normal approximations of the form 

$$
\hat \theta_k \sim \cN(\theta_k,\sigma^2_k) \quad \text{with either} \quad \begin{cases} 
\theta_k = \theta^*  \quad &\text{(fixed-effects),}
    \\
\theta_k \sim \cN(\theta^*,\tau^2)  \quad &\text{(random-effects)}.
\end{cases}
$$
The fixed-effects model (FE, $\theta_k=\theta^*$) assumes that all studies are measuring the same underlying effect. In contrast, the random-effects model (RE, $\theta_k \sim \cN(\theta^*,\tau^2)$) assumes that the reported effects are normally distributed around the effect of interest. The variance $\tau^2$ represents this between-study variability. %It can account for heterogeneity in study populations or study effects.

\paragraph{Estimation.} Usual estimators of $\theta^*$ take the form of inverse-variance weighting (IVW) estimators, that is
\begin{equation}
    \hat\theta^* = \sum_{k=1}^K \omega_k \hat\theta_k \quad \text{with} \quad \sum_{k=1}^K \omega_k = 1 \quad \text{and} \quad \omega_k \propto \frac{1}{\hat\sigma_k^2+\hat\tau^2}. \label{eq:ivw} 
\end{equation}
The between-study variance estimator $\hat \tau$ is set to $0$ for the FE model. For the RE model, it can be estimated by a number of methods, such as the DerSimonian and Laird estimator \citep{dersimonian1986meta} or the restricted maximum likelihood method \citep{viechtbauer2005bias}, just to cite a few.

\paragraph{Causal limitations.} 
The causal interpretation of a conventional meta-analysis estimates is often left implicit. In a meta-analysis of randomized trials, each study may identify a causal effect for its own study population, but the weighted average of these study-specific causal effects does not automatically equal the causal effect of treatment in a clearly defined target population. For the IVW estimator \eqref{eq:ivw}, these weights are primarily determined by statistical precision rather than by clinical relevance of the populations represented, so that the resulting estimand does not necessarily equal an average treatment effect in any actual population. This is particularly true for non-collapsible contrasts, which have the property that the total contrast associated with a population is not equal to an average of the stratified contrasts computed on sub-populations. We refer to \cite{berenfeld2025causal} for a thorough discussion on that matter, and to \cite{campbell2026hidden} for a discussion specific to odds-ratios.

\subsection{Network meta-analysis}

Network meta-analysis (NMA) extends pairwise meta-analysis to the comparison of more than two treatments. It combines direct evidence from studies that compare the treatments of interest within the same RCT with indirect evidence obtained through one or more common comparators. For example, trials comparing treatment ($a$) with treatment ($c$) together with trials comparing treatment ($b$) with treatment ($c$) may be used to estimate the treatment effect of ($a$) relative to ($b$), even when no trial comparing ($a$) and ($b$) directly.

We let $\cA = \{0,\dots, N\}$ be the set of treatment of interest. Like in the pairwise meta-analysis section, we assume that each study $k$ publishes the counts $n_k^{ay}$ of patients with treatment $A=a$ and outcome $Y=y$. However, unlike in the previous setting, each study only sees a subset $\cA_k \subset \cA$ of treatment, so that $n_k^a = 0$ for $a \notin \cA_k$. The data thus take the form of Table~\ref{tab:nma}.

\begin{table}[htbp]
\centering
\scalebox{0.85}{%
\begin{tikzpicture}[font=\sffamily]
% =====================================================
% Trial 1
% =====================================================
\begin{scope}[shift={(0,0)}]
\node[anchor=west] at (0,1.8) {%
\begin{tabular}{c|cc}
\textbf{Study 1} & $Y=1$ & $Y=0$ \\ \hline
\rowcolor{study1!20} $A = a$ & $n_1^{a1}$ & $n_1^{a0}$ \\
\rowcolor{study1!20} $A = b$ & $n_1^{b1}$ & $n_1^{b0}$ \\
\rowcolor{study1!20} $A = d$ & $n_1^{d1}$ & $n_1^{d0}$ \\
\end{tabular}};
\end{scope}
% =====================================================
% Trial 2
% =====================================================
\begin{scope}[shift={(5.5,0)}]
\node[anchor=west] at (0,1.8) {%
\begin{tabular}{c|cc}
\textbf{Study 2} & $Y=1$ & $Y=0$ \\ \hline
\rowcolor{study2!20} $A = a$ & $n_2^{a1}$ & $n_2^{a0}$ \\
\rowcolor{study2!20} $A = c$ & $n_2^{c1}$ & $n_2^{c0}$ \\
\rowcolor{study2!20} $A = e$ & $n_2^{e1}$ & $n_2^{e0}$ \\
\end{tabular}};
\end{scope}
% =====================================================
% Trial 3
% =====================================================
\begin{scope}[shift={(11,0)}]
\node[anchor=west] at (0,1.8) {%
\begin{tabular}{c|cc}
\textbf{Study 3} & $Y=1$ & $Y=0$ \\ \hline
\rowcolor{study3!20} $A = a$ & $n_3^{a1}$ & $n_3^{a0}$ \\
\rowcolor{study3!20} $A = b$ & $n_3^{b1}$ & $n_3^{b0}$ \\
\rowcolor{study3!20} $A = c$ & $n_3^{c1}$ & $n_3^{c0}$ \\
\end{tabular}};
\end{scope}
\end{tikzpicture}
}%
\caption{Arm-level outcome counts for each of the three trials in an hypothetic NMA.}
\label{tab:nma}
\end{table}

Network meta-analysis is often represented using a graph in which nodes denote treatments and edges represent studies that directly compared two treatments, see Figure~\ref{fig:network_meta_graph2} below. This formulation naturally accommodates multiple edges between the same pair of nodes, each corresponding to an independent study of that comparison. As a special case, a graph with only two nodes and multiple edges recovers the standard meta-analysis setting described in the previous section.

\begin{figure}[h]
  \centering
  %\begin{minipage}{0.45\textwidth}
    \centering
\begin{tikzpicture}[scale=0.6, 
  every node/.style={circle, draw, minimum size=1cm, font=\bfseries},
  every edge/.style={draw, thick}
]
  % Nodes
  \node (a) at (3, 4)   {$a$};
  \node (b) at (0, 2)   {$b$};
  \node (c) at (6, 2)   {$c$};
  \node (d) at (1.5, 0) {$d$};
  \node (e) at (4.5, 0) {$e$};
  % Study 1 (red): A1, A2, A4
  \draw[study1, thick, line width=1.5pt] (a) to[bend left=12]  (b); 
  \draw[study1, thick, line width=1.5pt] (a) -- (d);
  \draw[study1, thick, line width=1.5pt] (b) -- (d);
  % Study 2 (blue): A1, A3, A4, A5
  \draw[study2, thick, line width=1.5pt] (a) to[bend left=12]  (c); 
  \draw[study2, thick, line width=1.5pt] (c) -- (e);
   \draw[study2, thick, line width=1.5pt] (a) -- (e);
  % Study 3 (green): A1, A2, A3
  \draw[study3, thick, line width=1.5pt] (a) to[bend right=12] (b);
  \draw[study3, thick, line width=1.5pt] (a) to[bend right=12] (c);
  \draw[study3, thick, line width=1.5pt] (b) -- (c);
\end{tikzpicture}

  \caption{An illustration of the NMA setting of Table~\ref{tab:nma}. The treatment nodes are connected by edges coloured by study \textcolor{study1}{$k=1$}, \textcolor{study2}{$k=2$}, and \textcolor{study3}{$k=3$}. %Multiple edges between the same pair of nodes (e.g.\ $A_1$--$A_2$, $A_1$--$A_3$) indicate repeated comparisons across different studies. 
  The absence of an edge between $(d)$ and $(e)$ indicates that no study directly comparing these treatments is available.}
  \label{fig:network_meta_graph2}
\end{figure}
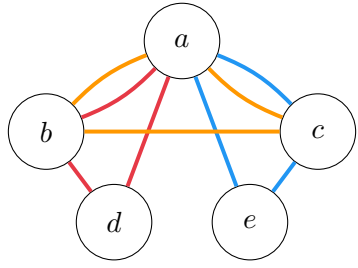

Similarly as before, we let $\theta_k^{ab}$ (resp. $\hat\theta_k^{ab}$) be the true (resp. reported) contrast between treatment $a$ and $b$ in study $k$, and $\psi^{a}_k$ (resp $\hat \psi^{a}_k$) be the true (resp. reported) absolute effect of treatment $a$ in study $k$. We thus have 
$$
\theta_{k}^{ab} = h(\psi_{k}^{a})-h(\psi_{k}^{b}) \quad \text{and} \quad \hat \theta_{k}^{ab} = h(\hat\psi_{k}^{a})-h(\hat\psi_{k}^{b}) \quad\text{with}\quad \hat\psi_k^c := \frac{n_k^{c1}}{n_k^c} \quad \text{for} \quad c\in \cA_k.
$$
where we recall that $h$ is the link function defined in Section~\ref{sec:Reminders}.

Two broad likelihood frameworks have been proposed for NMA, differing in the level at which the data are modelled.

 \paragraph{Contrast-likelihood (CL).} CL approaches directly model the distribution of observed contrasts $\hat\theta_{k}^{ab}$ between treatments $a$ and $b$ in each study $k$. It corresponds to a model on the \emph{edges} of the network graph (as in Figure~\ref{fig:network_meta_graph2}). More precisely, one assumes that there exists $d^{ab}$ such that, for $a,b \in \cA_k$:
\[
  \hat\theta_{k}^{ab} \sim \cN(\theta_k^{ab}, (\sigma_k^{ab})^2),
  \quad \text{with either} \quad  \begin{cases}
    \theta_k^{ab} = d^{ab} \quad &\text{(fixed-effects)}, \\ 
    \theta_k^{ab} \sim \cN(d^{ab},\tau^2) \quad &\text{(random-effects)},
  \end{cases}
\]
where $(\sigma_{k}^{ab})^2$ is the known sampling variance of $\hat\theta_{k}^{ab}$ and $\tau$ is a between-study variance. 
A key assumption in network meta-analysis is the \emph{transitivity} assumption:
\begin{equation}\label{eq:transitivity}
  d^{ac} = d^{ab} + d^{bc}, \qquad \forall\, a,b,c \in \mathcal{A},
\end{equation}
which enables leveraging  indirect comparisons to get an estimate of the relative effect between two treatments. This assumption is equivalent to the existence of real numbers $(d^a)_{a \in \mathcal{A}}$ satisfying $d^{ab} := d^a - d^b$, with $d^1 := 0$ for a reference treatment $1$. Stacking the observed contrasts into a vector $\hat{\boldsymbol{\theta}}$ and the treatment parameters into $\mathbf{d} = (d^a)_{a \in \mathcal{A}}$, the
model rewrites in matrix form as
\begin{equation}\label{eq:cl_model}
  \hat{\boldsymbol{\theta}} = Z\mathbf{d} + \boldsymbol{\varepsilon}
  + \boldsymbol{\eta} 
  \qquad
  \boldsymbol{\varepsilon} \sim \mathcal{N}({0}, \Sigma),
  \qquad
  \boldsymbol{\eta} \sim \mathcal{N}({0}, \tau^2 V),
\end{equation}
where $\Sigma$ is the known sampling variance matrix and $Z$ is the \emph{edge-node incidence matrix}, encoding how treatments are linked through studies. The heterogeneity covariance has block-diagonal structure $V = \mathrm{diag}(V_1,\ldots,V_K)$, where each block $V_k \in \mathbb{R}^{v_k \times v_k}$ has ones on the diagonal and $1/2$ off-diagonal, see e.g. \cite{white2019armcontrast}. Under the FE model ($\tau=0$), one has an explicit solution of the maximum likelihood estimator of $\bd$:
$$
  \hat{\mathbf{d}}^{\mathrm{FE}}
  = \bigl(Z^\top \Sigma^{-1} Z\bigr)^{+} Z^\top \Sigma^{-1} \hat{\boldsymbol{\theta}}.
$$
Under the RE model, as in pairwise meta-analysis, $\tau^2$ is first estimated using the methods of Section~\ref{sub_sec:pairwise_meta} (see the Estimation paragraph) and then plugged into the MLE to find
$$
 \hat{\mathbf{d}}^{\mathrm{RE}}
  = \bigl(Z^\top (\Sigma+\hat \tau^2 V)^{-1} Z\bigr)^{+} Z^\top (\Sigma+\hat \tau^2 V)^{-1} \hat{\boldsymbol{\theta}}.
$$
One can check that one retrieves the IVW estimator of \eqref{eq:ivw} in the pairwise case (i.e. $\cA = \cA_k = \{0,1\}$ for all $k$). 
\begin{rem}
    The product $Z \hat{\mathbf{d}}^{\mathrm{FE}}$ can be interpreted as the $\Sigma^{-1}$-weighted projection of $\hat{\boldsymbol{\theta}}$ onto the subspace of transitive contrasts (see \emph{e.g.}~\cite{rucker2012network}), thereby assigning greater influence to pairwise estimates with smaller variance.
\end{rem}

  \paragraph{Arm-likelihood (AL).} AL approaches, by contrast, model the observed outcome $\hat\psi_{k}^a$ of each arm $a$ in study $k$ individually. It corresponds to a model on the \emph{nodes} of the network graph. The observed outcome in arm $a$ of study $k$ is assumed to follow 
 $$n_k^{a1} \mid \psi_k^a \sim \mathrm{Bin}(n_k^a, \psi_k^a),$$ 
 independently across arms and studies, where $\psi_k^a \in (0,1)$ is the unknown study-specific arm-effects. 

At the second stage, since $\psi_k^a \in (0,1)$, we model it on an
unconstrained scale via the link function $h$, setting
$\mu_k^a := h(\psi_k^a)$.  Two parametrizations are considered: an \emph{arm-based} (AB) parametrization through arm-level parameters, and a \emph{contrast-based} (CB) parametrization through treatment contrast:
\begin{itemize}
  \item \emph{Arm-based (AB):} each treatment is assigned an absolute
  parameter $\mu^a$, and assume that
  \[
    \mu_k^a = \mu^a + \eta_k^a,
    \qquad \eta_k^a \sim \mathcal{N}(0,\, \tau^2),
  \]
  where $\eta_k^a$ is a random effect. 

  \item \emph{Contrast-based (CB):} arm-level effects are expressed through
  differences $d^a - d^{a_k^*}$ relative to a study-specific reference
  $a_k^* \in \mathcal{A}_k$,
  \[
    \mu_k^a = \alpha_k + d^a - d^{a_k^*} + \eta_k^a,
    \qquad \eta_k^a \sim \mathcal{N}(0,\, \tau^2),
  \]
  where $\alpha_k$ is the baseline risk for study $k$ on the transformed scale. A further extension of the contrast-based parametrization allows random baselines $\alpha_k \sim \mathcal{N}(\alpha, \sigma^2)$, independently of the contrast random effects~\citep{white2019armcontrast}.
\end{itemize}
In both cases, the FE model corresponds to $\tau = 0$ and the RE model allows $\tau^2 > 0$. The parameters $(\boldsymbol{\mu}, \tau)$ (resp. $(\boldsymbol{\alpha},\boldsymbol{d},\tau)$) are usually jointly estimated by maximizing the likelihood numerically. A closed-form MLE exists only for the arm-based FE model ($\tau=0$), in which case
\begin{equation}\label{eq:FE_MLE}
    \hat\mu^{a} = h\!\left(\frac{\sum_{k\in \cK^a} n_k^{a1}}{\sum_{k\in\cK^a} n_k^a}\right),  
\end{equation}

\paragraph{Causal limitations.}

The estimands considered in  this section are again defined purely in statistical terms, with no explicit reference to a target population or intervention. This problem is especially important for indirect comparisons. If trials of (A) versus (C) and trials of (B) versus (C) involve different populations, the indirect estimate of (A) versus (B) is causally interpretable only if the relevant effects can be transported to a common population. For instance, both \cite{jansen2013network} and \citet[Chap. 11]{chandler2019cochrane} states that indirect comparisons may be biased when effect modifiers are unevenly distributed across studies.

A related debate is the CB vs AB controversy in network meta-analysis \citep{white2019armcontrast}. The latter has been criticised for \emph{breaking the randomization} \citep{dias2016absolute} in the sense that the estimated contrast between two treatments may reflect differences in baseline risks or settings between two studies rather than an actual treatment effect. The causal concern is therefore different in the two approaches: in AB models, causal interpretation may require strong assumptions about the exchangeability of  risks across studies; in CB models, the within-study comparisons are protected by randomization, but indirect comparisons still require transitivity, which in turn entails strong assumptions about the distribution of treatment effect modifiers. \medskip

We now depart from the classical statistical view and instead frame these quantities through a causal lens, recasting them as population-level causal effects. This point of view offers a principled approach to address heterogeneity across studies by explicitly modeling it as a consequence of the underlying causal mechanisms.

\section{Causal meta-analysis: the pairwise case} \label{sec:causalpair}

This section begins by introducing a causal framework for pairwise meta-analysis following~\cite{berenfeld2025causal}, and examines settings with varying degrees of heterogeneity: the homogeneous, no-center-effect setting, where all studies share the same population and the study has no direct effect on the outcome; the heterogeneous, no-center-effect setting, where populations may differ across studies but there is still no direct study effect; the homogeneous-population, center-effect setting, where the population is fixed but a direct study effect is present; and the most general setting, where both population heterogeneity and a direct center effect are allowed.

\subsection{A causal framework for meta-analysis} 

 We represent an individual  by a realization of a random tuple $(H, X, A, Y)$, where $H \in [K]$ denotes the study indicator, $A \in \{0,1\}$ denotes the treatment indicator, $X$ is a vector of covariates, and $Y$ is the outcome. We start with a very natural positivity assumption that ensures that every study have actually enrolled participants.
\begin{ass}[Study positivity] $\bbP(H=k) > 0$ for all $k \in [K]$. \label{ass:studypos}  
\end{ass}

 Following the potential outcome framework of~\cite{rubin1976inference}, we posit the existence of two random variables, termed counterfactual variables $Y^0$ and $Y^1$, corresponding to the outcomes an individual  would have experienced had they received treatment $0$ or $1$, respectively.  
\begin{ass}[SUTVA] $Y=AY^1+(1-A)Y^0$. \label{ass:sutva}  
 \end{ass}

This first assumption (SUTVA, for Stable Unit Treatment Value Assumption) relates the observed outcome $Y$ to its potential outcomes $Y^a$, and states that each participant's outcome depends only on their own treatment assignment, and that treatment is administered in the same way for all participants.

To compare the effect of two treatments, one is usually interested in computing
\begin{align}\label{eq:causal_contrast}
    \Phi\big(\mathbb{E}[Y^1],\mathbb{E}[Y^0]\big),
\end{align}
for some contrast function $\Phi$. For example, choosing $\Phi(x,y)=x-y$ yields the \emph{Average Treatment Effect} (ATE) with the risk difference (RD),
\[
\text{ATE}=\mathbb{E}[Y^1-Y^0],
\]
whereas $\Phi(x,y)=x/y$ will lead to the risk-ratio (RR).  In causal inference, the goal is to estimate this quantity from data alone, as for each unit only one of the two potential outcomes $Y^1$ and $Y^0$ is observed, never both. This is known as the \emph{fundamental problem of causal inference}. To achieve identification, a common assumption on the data generating process is that treatment assignment is independent of the potential outcomes, as is guaranteed by design in a \emph{Randomized Controlled Trial} (RCT). In the meta-analytic setting, we assume that each study conducted an RCT, formally:
\begin{ass}[Collection of RCTs]
$A \indep Y^0, Y^1 \mid H$.
\label{ass:rcts}
\end{ass}

Let $P_k$ be the distribution of the covariates in study $k$ and define the absolute effect under treatment $A=a$ in study $k$ as
\begin{align}\label{eq:psi_ka}
    \psi_k^a := \bbE[Y^a\mid H=k] = \bbE_{P_k}\big[\underbrace{ \bbE[Y^a\mid H=k,X=x]}_{:= \mu_k(a,x)}\big],
\end{align}
where the function $\mu_k$ is termed the per-study  \emph{baseline/response function.}
Similarly, let the contrast in study $k$ be defined as $\theta_k := \Phi(\psi_k^1,\psi_k^0)$ for a contrast function $\Phi$. Following~\cite{berenfeld2025causal}, we observe that $\theta_k$ can be written only as a function of $\mu_k$ and $P_k$, and we write:
$$
\theta_k := \theta(\mu_k,P_k).
$$

We say in this work that an estimand $\theta^\star$ is causal if it can be put on the form $\theta^* = \theta(\mu^*,P^*)$ for some specific function $\mu^*$ and target population $P^*$. 
%For a more formal discussion and definition see Appendix~\ref{app:causa_inter_dfn}.
For the remainder of the paper, we investigate under which conditions on the $\mu_k$'s and the $P_k$'s such causal estimands can be targeted using only the aggregated-data 
$$
\{n_k^{ay}~:~k\in[K], a\in\{0,1\}, y\in\{0,1\}\},
$$
what the corresponding estimators are, and what their properties are. Namely we consider four settings: one where both the $P_k$'s and $\mu_k$'s are fixed to a common value (Section~\ref{sec:pwhomo}), one where only the $P_k$'s are allowed to differ (Section~\ref{sec:pwpop}), one where only the $\mu_k$'s are allowed to differ (Section~\ref{sec:pwcenter}), and one where both can differ (Section~\ref{sec:pwhetero}).

\subsection{Causal meta-analysis with homogeneous population} \label{sec:pwhomo}
This section addresses the case of a homogeneous population, in which patients across all studies are drawn from a single underlying population. Formally, in addition to Assumptions ~\ref{ass:sutva}-\ref{ass:studypos} we assume that

 \begin{ass}[Homogeneous population] $H \indep X$. \label{ass:homop} 
 \end{ass}
Additionally, we start by assuming that there is no direct effect from the trial on the outcome.
This assumption is often coined
\emph{exchangeability in mean, no center-effect} ~\citep{robertson2021center, khellaf2025federatedO}, or
sometimes \emph{weak response consistency} ~\citep{sobel2017causal}. 

 \begin{ass}[No-center effect] $H \indep Y^1,Y^0 \mid X$. \label{ass:nocenter} 
 \end{ass}

This assumption implies the transportability of the conditional mean of the potential outcomes across trials, namely $\mathbb{E}[Y^a\mid X,H=k] = \mathbb{E}[Y^a\mid X,H=\ell]$ for all $k,\ell \in [K]$. In other words the outcome model given treatment and covariates is assumed to be the same in every trial.

Finally, we assume each treatment is given out at least some of the time — i.e., never with zero probability — but we do not assume that this probability is constant across studies; it may vary from one study to another.
\begin{ass}[Treatment positivity] \label{ass:postrt} $\bbP(A=a) > 0$ for $a \in \{0,1\}$.
\end{ass}

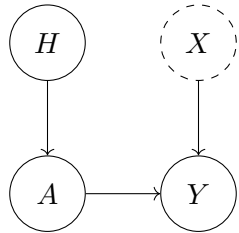
\begin{figure}[h]
\centering
\begin{tikzpicture}[
    node distance=2cm,
    every node/.style={circle, minimum size=1cm},
    observed/.style={draw, solid},
    unobserved/.style={draw, dashed}
]
    \node[observed]   (H) at (0,2) {$H$};
    \node[unobserved] (X) at (2,2) {$X$};
    \node[observed]   (A) at (0,0) {$A$};
    \node[observed]   (Y) at (2,0) {$Y$};

    \draw[->] (H) -- (A);
    \draw[->] (X) -- (Y);
    \draw[->] (A) -- (Y);
\end{tikzpicture}
\caption{The directed acyclic graph (DAG) representing the causal structure of the setting with homogeneous populations and no center effect. The variable $X$ is not  observed.}
\label{fig:homogeneous_pop&resp}
\end{figure}

Under Assumptions~\ref{ass:sutva}-\ref{ass:postrt} (see the corresponding DAG in Figure \ref{fig:homogeneous_pop&resp}) let  $P_0$ denote the common covariate distribution  and  $\mu_0$ denote the common response function. For $P^* = P_0$ and $\mu^* = \mu_0$, one obtains the identification formula 
\begin{equation}\label{eq:identif_homogen&no_center}
    \psi^{*a} = \bbE_{P_0}[\mu_0(a,X)] = \bbE_{P_0}[\bbE[Y^a\mid X]] = \bbE[Y^a] = \bbE[Y \mid A=a],
\end{equation}
and naturally 
leads to the estimator 
$$
\hat\psi^a := \frac{n^{a1}}{n^a}.
$$
This estimator targets the absolute causal effect across a population distributed according to  $P_0$ and with response function equal to $\mu_0$. 

\subsection{Causal pairwise meta-analysis with heterogeneous population}\label{sec:pwpop}
\begin{figure}[h]
\centering
\begin{tikzpicture}[
    node distance=2cm,
    every node/.style={circle, minimum size=1cm},
    observed/.style={draw, solid},
    unobserved/.style={draw, dashed}
]
    \node[observed]   (H) at (0,2) {$H$};
    \node[unobserved] (X) at (2,2) {$X$};
    \node[observed]   (A) at (0,0) {$A$};
    \node[observed]   (Y) at (2,0) {$Y$};

    \draw[->] (H) -- (X);
    \draw[->] (H) -- (A);
    \draw[->] (X) -- (Y);
    \draw[->] (A) -- (Y);
\end{tikzpicture}
\caption{The directed acyclic graph (DAG) representing the causal structure of the setting with heterogeneous populations and no center-effect. The variable $X$ is not observed.}
\label{fig:heterogeneous&no_centerE}
\end{figure}
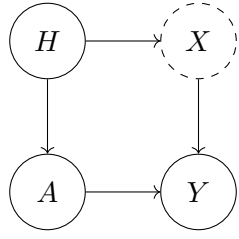

We now relax the homogeneous population assumption (Assumption~\ref{ass:homop}) by allowing each study to arise from a distinct population. Indeed, even when clinical trials share the same inclusion criteria, it is natural to expect differences in the distribution of covariates across studies. In our framework, these covariates remain unobserved.
This setting is represented by the DAG in Figure~\ref{fig:heterogeneous&no_centerE}, through the addition of an arrow from $H$ to $X$\footnote{The choice of the direction of the arrow between $X$ and $H$ can be up to debate. When drawn from $X$ to $H$, the variable $H$ has the meaning of an inclusion variable: based on their covariate, the patient is included or not in study $H$. When drawn from $H$ to $X$, the variable $H$ has the meaning of a context variable: if a patient belongs to study $H$, then the distribution of its covariate is the one of the patients if this study. Note that both DAGs are Markov equivalent and thus represents the same set of data distributions.}.
In this section, we keep Assumptions~\ref{ass:sutva}-\ref{ass:studypos} together with the no-center effect assumption (Assumption~\ref{ass:nocenter}), since there remains no direct arrow from $H$ to $Y$. Thus, while covariate distributions may vary between studies, the relationship between the potential outcomes and the covariates is assumed to be the same in every study.
The treatment positivity assumption is however replaced by Assumption~\ref{ass:pos2} below. The material of this section is drawn from \cite{berenfeld2025causal}, to which we refer the reader for further details.
\begin{ass}[Treatment positivity II]  \label{ass:pos2} $\bbP(A=a \mid H = k) > 0$ for all $a \in \{0,1\}$ and $k \in [K]$.
\end{ass}

We denote by $P_k$ the covariate distribution in study $k$. The objective is to estimate a treatment effect in a target population $P^\star$, defined as a mixture of the study-specific populations. Formally, given non-negative weights $\alpha_k$ such that $\sum_{k=1}^K \alpha_k = 1$, \cite{berenfeld2025causal} consider target populations of the form
$$
P^\star = \sum_{k=1}^K \alpha_k P_k.
$$

The choice of weights determines the target population and therefore the estimand of interest. Natural choices include weighting studies according to their sample sizes, $\alpha_k=n_k/n$, which targets the pooled trial population, or assigning equal weights, $\alpha_k=1/K$, which gives the same importance to each study. Other choices may be more appropriate in practice. For example, if the objective is to assess the effect of a treatment in a particular market or healthcare setting, greater weight can be assigned to studies that are most representative of that population. Once a target population has been specified, the treatment effect can be estimated. The resulting estimators generally differ from classical fixed-effect and random-effects meta-analytic estimators and can be viewed as arm-based estimators.

More precisely, from the no-center effect assumption, we have $\mu_k = \mu_0$ for all $k$, which, under Assumptions~\ref{ass:studypos}-\ref{ass:rcts}, \ref{ass:nocenter} and \ref{ass:pos2} yields the identification formula
\begin{align*}
\psi^{*a} &= \bbE_{P^\star}\bbE[Y^a|X] \overset{(*)}{=} \bbE_{P^\star}[\mu_0(a,X)]=\sum_{k=1}^K \alpha_k \bbE_{P_k}\left[\mu_0(a,X)\right] =  \sum_{k=1}^K \alpha_k \bbE_{P_k}\left[\mu_k(a,X)\right]\\ &= \sum_{k=1}^K \alpha_k \bbE\left[Y^a\mid H=k\right] = \sum_{k=1}^K \alpha_k \bbE\left[Y\mid H=k, A=a\right],
\end{align*}
where $(\ast)$ follows from Assumption~\ref{ass:nocenter}. 
This identification formula suggests the estimator
\begin{equation}\label{eq:populationMixture_estimator}
  \hat \psi^a := \sum_{k=1}^K \hat\alpha_k \frac{n_k^{a1}}{n_k^a}= \sum_{k=1}^K \hat\alpha_k \hat{\psi}_k^a.  
\end{equation}

\begin{prp}\label{prp:consistency_pop}  Under Assumptions~\ref{ass:studypos}-\ref{ass:rcts}, \ref{ass:nocenter} and \ref{ass:pos2}, and given consistent estimators $\hat\alpha_k$ of $\alpha_k^*$, the absolute effects estimators $\hat\psi^a$ are consistent estimators of $\psi^{*a}$. 
\end{prp}

The proof of the latter restult is straightforward. We can also get formulas for the variance. When $\hat\alpha_k = \alpha_k^*$ are set to deterministic weights, we have the following result. We refer to \cite{berenfeld2025causal} for the variance formulas with weights $\hat\alpha_k = n_k/n$.

\begin{prp} \label{prp:variance_pop} Assumptions~\ref{ass:studypos}-\ref{ass:rcts}, \ref{ass:nocenter} and \ref{ass:pos2}, and if $\hat\alpha_k = \alpha_k^*$, the estimator $\boldsymbol{\hat\psi} := (\hat\psi^1,\hat\psi^0)$ is asymptotically normal with $\sqrt{n}(\boldsymbol{\hat \psi}-\boldsymbol{\psi}^*) \to \cN(0,\Sigma)$. A consistent estimator of $\Sigma$ is given by
$$
\hat\Sigma^{aa} = n \sum_{k=1}^K \alpha_k^{*2} \frac{\hat\psi_k^a(1-\hat\psi_k^a)}{n_k^a}.
$$
and $\hat \Sigma^{ab}=0$ for $a\neq b$.
\end{prp}

The proof of this result is also straightforward but can be found in Appendix~\ref{app:variance_pop} for completeness. In this setting, if $\alpha_k^* \asymp 1/K$, the resulting variance is of order $(K n \eta_K \ve)^{-1}$ where $\eta_K = \min_k \bbP(H=k)$ and $\ve = \min_{a,k} \bbP(A=a \mid H=k)$. 

The treatment relative effects are obtained $\theta := \Phi(\psi^1,\psi^0)$ for a contrast function $\Phi$.
The key distinction from fixed/random effects approaches lies in the ordering of the contrast and averaging operations: while the standard setting computes contrasts before averaging, the causal approach reverses this order by averaging first and then taking the contrast. \cite{berenfeld2025causal} further illustrate this distinction on the risk-ratio. With two studies and equal weights $\alpha_1 = \alpha_2 = 1/2$, and letting $\psi_{k}^{a} = \mathbb{E}[Y^a \mid H = k]$ denote the study-specific expected potential outcomes, log-scale aggregation yields a random-effects estimand of the form
\[
\theta^{\rm RE}_{\rm RR} = \sqrt{\frac{\psi_1^1\psi_2^1}{\psi_1^0\psi_2^0}},
\]
a geometric mean of study-specific risk ratios, while the causal estimand corresponds to a genuine contrast of averages,
\[
\theta^{\rm causal}_{\rm RR} = \frac{\psi_1^1+\psi_2^1}{\psi_1^0+\psi_2^0}.
\]
The two admit no natural ordering. In particular, $\theta^{\rm RE}_{\rm RR}$ is much more sensitive to small values of $\psi_k^0$, the baseline risk (thus more likely to be driven by a single study), and the random-effects estimand has no representation as a contrast of an average, hence no causal interpretation.
They further show that classical and causal meta-analytic estimands may diverge when between-study heterogeneity is substantial. In many realistic settings, however, these discrepancies remain limited, which is reassuring from a public health perspective.

\subsection{Causal pairwise meta-analysis with center-effect and homogeneous population}\label{sec:pwcenter}
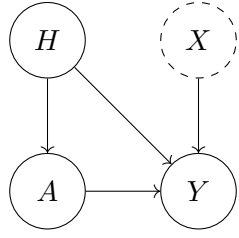
\begin{figure}[h]
\centering
\begin{tikzpicture}[
    node distance=2cm,
    every node/.style={circle, minimum size=1cm},
    observed/.style={draw, solid},
    unobserved/.style={draw, dashed}
]
    \node[observed]   (H) at (0,2) {$H$};
    \node[unobserved] (X) at (2,2) {$X$};
    \node[observed]   (A) at (0,0) {$A$};
    \node[observed]   (Y) at (2,0) {$Y$};

    \draw[->] (H) -- (A);
    \draw[->] (H) -- (Y);
    \draw[->] (X) -- (Y);
    \draw[->] (A) -- (Y);
\end{tikzpicture}
\caption{The directed acyclic graph (DAG) representing the causal structure of the setting with homogeneous populations but with a center-effect (arrow from $H$ to $Y$). The variable $X$ is not observed.}
\label{fig:center_effect&homogeneousPop}
\end{figure}

This section considers a setting in which heterogeneity arises from center-specific effects, such as variation in outcome measurement,  while the underlying population remains homogeneous across sites. 
We assume homogeneous populations as in Assumption~\ref{ass:homop} but, contrary to the previous section, we do not assume that the baseline/response function $\mu_k$ is the same across centers. We thus work under assumptions~\ref{ass:studypos}-\ref{ass:homop} and \ref{ass:pos2}, and refer to Figure~\ref{fig:center_effect&homogeneousPop} for the corresponding DAG. 

Here, our causal estimand of interest is defined over the population that is common to all studies, but with a response function that is an average of the individual studies' response functions. Specifically, let
$$
\mu^*(a,X) = \sum_{k=1}^K \beta_k \mu_k(a,X).
$$
and observe that, on a population $(X,A , Y^0, Y^1)$ with $P_X=P_k=P_0$ for all $k$ and with response function $\mu^\star(a,x)=\mathbb{E}[Y^a|X=x]$, one has
\begin{align*}
\psi^{*a} = &\bbE_{P_0}\bbE[Y^a|X] = \bbE_{P_0}[\mu^\star(a,X)]=\sum_{k=1}^K \beta_k \bbE_{P_k}\left[\mu_k(a,X)\right] =  \sum_{k=1}^K \beta_k \bbE\left[Y^a\mid H=k\right]\\
=& \sum_{k=1}^K \beta_k \bbE\left[Y\mid H=k, A=a\right].
\end{align*}
Consequently, given estimators $\hat{\beta}_k$ of the weights $\beta_k$, the estimator
$$
\hat{\psi}^a := \sum_{k=1}^K \hat{\beta}_k\, \frac{n_k^{a1}}{n_k^a}= \sum_{k=1}^K \hat{\beta}_k\, \hat{\psi}_k^a.
$$

The weights $\beta_k$ play a different role here compared to the weights $\alpha_k$ introduced in the previous section. While the latter were defined over populations and could reflect, for instance, the relative size of each subgroup,  the weights $\beta_k$ can be viewed as a measure of \emph{reliability} across studies. In this setting, a natural criterion for assigning a larger weight to a particular study is the degree of \emph{confidence} one places in its results --- whether due to more rigorous experimental designs, lower risk of bias, or greater methodological transparency. Results concerning consistency and asymptotic normality are the same as in Section~\ref{sec:pwpop} (Propositions~\ref{prp:consistency_pop} and \ref{prp:variance_pop}) but under Assumptions~\ref{ass:studypos}-\ref{ass:homop} and \ref{ass:pos2}.

\begin{rem}
The estimator introduced in this section is equivalent to the one proposed in~\cite{zhang2026causal}. However, unlike the present work, the authors therein consider a setting consistent with the graph in Figure~\ref{fig:center_effect&homogeneousPop} without distinguishing between the effects of $H$ and $X$ on $Y$.
\end{rem}

\subsection{Causal pairwise meta-analysis with center-effect and heterogeneous population}\label{sec:pwhetero}

The goal of this section is to study a fully heterogeneous setting in which both the populations and the baseline functions are allowed to differ;  formally, we drop Assumptions~\ref{ass:homop} and~\ref{ass:nocenter}. As in the previous scenarios, each quantity $\theta_k = \theta(\mu_k, P_k)$ is causal for every $k$, and the objective is again to show that the natural arm-based aggregation estimator targets a well-defined absolute causal effect, yielding a well-defined causal contrast $\theta(\mu^\star, P^\star)$. Informally, the idea is to take $\mu^\star$ to be the the average of the per-study baseline/response functions $\mu_k$ and $P^\star$ to be the average of the covariate distributions $P_k$. To make this precise, this section introduces a hierarchical model and views the pairs $(\mu_k,P_k)$ as random samples from $\Pi$; the means $\mu^\star$ and $P^\star$ are then defined as the expectations under this law $\Pi.$ The crucial assumption is that $\mu_k \indep P_k$, \emph{i.e.,} the law $\Pi$ factorizes as a product between a law on $\mu_k$ and a law on $P_k$, which, informally,  allows study-effects but requires them to be
uncorrelated with \emph{recruitment}: knowing a study sampled some
covariate region tells you nothing about its outcome mechanism.

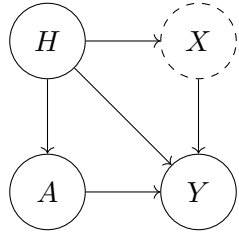
\begin{figure}[h!]
\centering
\begin{tikzpicture}[
    node distance=2cm,
    every node/.style={circle, minimum size=1cm},
    observed/.style={draw, solid},
    unobserved/.style={draw, dashed}
]
    \node[observed]   (H) at (0,2) {$H$};
    \node[unobserved] (X) at (2,2) {$X$};
    \node[observed]   (A) at (0,0) {$A$};
    \node[observed]   (Y) at (2,0) {$Y$};

    \draw[->] (H) -- (X);
    \draw[->] (H) -- (A);
    \draw[->] (H) -- (Y);
    \draw[->] (X) -- (Y);
    \draw[->] (A) -- (Y);
\end{tikzpicture}
\caption{The directed acyclic graph representing the causal structure of the setting with heterogeneous populations and with center-effect (arrow from $H$ to $Y$). The variable $X$ is not observed.}
\label{fig:fullyHeterogeneous}
\end{figure}

\paragraph{Hierarchical model.} There is a law $\Pi$ on triples $(P,\mu,e)$, where  \begin{itemize}[noitemsep]
\item $P\in\mathcal{P}(X)$ is a covariate distribution;
\item $\mu: \{0,1\} \times \cX \to [0,1]$ is the response/baseline function;
\item $e\in[0,1]$ is the assignment probability, parametrizing the law of
$A\mid H$ as $\mathrm{Ber}(e)$.
\end{itemize}
Independently across studies $k=1,\dots,K$, we sample
\[
(P_k,\mu_k,e_k)\ \sim\ \Pi.
\]
Given $(P_k,\mu_k,e_k)_{k\in[K]}$, each patient's data is a realization of a random tuple $(H,X,A,Y)$ where:
\begin{enumerate}
    \item $H \sim \cP_H$ where $\cP_H$ is a given distribution over $[K]$;
    \item $X \mid H =k \sim P_k$;
    \item $A \mid H=k \sim \mathrm{Ber}(e_k)$;
    \item $Y^a \mid X=x,H=k \sim \mathrm{Ber}(\mu_k(a,x))$;
\end{enumerate}
and, similar to the previous sections, we assume that Assumption ~\ref{ass:sutva} holds, \emph{i.e.,} $Y=AY^1+(1-A)Y^0$. Treatment assignment uses only $H$ (and is independent of
$(X,Y^0,Y^1)$ given $H$), so within each study Assumption~\ref{ass:rcts} holds. The resulting DAG is represented in Figure~\ref{fig:fullyHeterogeneous}.

\paragraph{The targeted estimand.}
Define the \emph{mean baseline/response} and the \emph{mean population}
\begin{equation}
\mu^\star(a,x) := \E_\Pi[\mu_k(a,x)],
\qquad
P^\star := \E_\Pi[P_k],
\label{eq:mean_baseline}
\end{equation}
the latter being defined as: $P^\star(B):=\mathbb{E}_\Pi[P_k(B)]$ for a measurable set $B$. The  estimand targeted in this section is the causal quantity 
\begin{equation*}
\theta(\mu^\star,P^\star)
= \Phi\!\left(\psi^{*1},\psi^{*0}\right) \quad \text{with}\quad \psi^{*a} := \bbE_{P^\star}[\mu^\star(a,X)].
\end{equation*}

\begin{ass}\label{ass:response_cov_indep}
    Under $\Pi$, the outcome mechanism is independent of the covariate
distribution, 
that is,  $\mu_k\indep P_k$.
\end{ass}

\begin{rem}
Assumption~\ref{ass:response_cov_indep} allows study-effects but requires them to be
uncorrelated with \emph{recruitment}: knowing a study sampled some
covariate region tells you nothing about its outcome mechanism. It is
strictly weaker than Assumption~\ref{ass:nocenter} ($\mu_k=\mu_0$), the degenerate case
in which $\mu_k$ carries no randomness.  
\end{rem}

Note that for each $a \in \{0,1\}$,
\begin{equation}\label{eq:ident_hier}
\mathbb{E}_\Pi\big[\psi_k^a\big]
=\mathbb{E}_\Pi \Big[\mathbb{E}_{P_k}\big[\mu_k(a,X)\big]\Big]
\overset{(\ast)}{=}\int_\mathcal{X} \E_\Pi[\mu_k(a,x)]\diff\E_\Pi[P_k](x)
=\mathbb{E}_{P^\star} [\mu^\star(a,X)]
\end{equation}
where $(\ast)$ follows by the independence assumption, \emph{i.e.,} Assumption~\ref{ass:response_cov_indep}.
and, consequently, for any contrast $\Phi$,
\begin{equation*}
\Phi\big(\E_\Pi[\psi_k^1],\,\E_\Pi[\psi_k^0]\big)
\;= \; \Phi\!\left(\bbE_{P^\star}[\mu^\star(1,X)],\ \bbE_{P^\star}[\mu^\star(0,X)]\right)\;=\;
\theta(\mu^\star,P^\star).
\end{equation*}
This identification formula yields a natural estimator for $\theta^*$:
\[
\hat\psi^a:=\frac1K\sum_{k=1}^K\hat\psi_k^a = \frac1K\sum_{k=1}^K\frac{n_k^{a1}}{n_k^a}
\quad \text{and}\quad
\hat\theta = \Phi\big(\hat\psi^1,\hat\psi^0\big).
\]

In terms of asymptotic guarantees, this translates into a result that is joint in $K$ and $n$, meaning that both the number of studies and the number of patients are allowed to grow simultaneously, alongside a hierarchical analogue of the overlap condition ensuring that each treatment is assigned with non-negligible probability across studies and centers.

\begin{ass}[Uniform treatment positivity]  \label{ass:unif_trt}
    There exists $0<\varepsilon<1$ such that $e_k \in [\varepsilon, 1-\varepsilon]$ almost-surely.
\end{ass}

Under the latter assumption, we find the following result whose proof can be found in Appendix~\ref{app:consistency_hier}.

\begin{prp} \label{prp:consistency_hier} Let $\eta_K =\min_{k \in [K]}  \mathbb{P}(H=k)$. In the hierarchical model with Assumptions~\ref{ass:response_cov_indep} and \ref{ass:unif_trt}, 
    as $K \to \infty$ and $n\to\infty$
jointly with  $K \eta_K n \to \infty$,
the estimator $\boldsymbol{\hat \psi} = (\hat\psi^1,\hat\psi^0)$ is a consistent estimator of the target absolute treatment effects $\boldsymbol{\psi}^* = (\psi^{*1},\psi^{*0})$:
\[
\boldsymbol{\hat\psi} \;\xrightarrow{\ \mathrm{p}\ }\; \boldsymbol{\psi}^*.
\]
\end{prp}

To the price of a stronger condition on how $K,n \to\infty$, one can also get asymptotic normality for the absolute treatment effect estimator.

\begin{prp} \label{prp:variance_hetero_pairwise} In the hierarchical model with Assumptions~\ref{ass:response_cov_indep} and \ref{ass:unif_trt}, and in the regime where $n,K \to \infty$ and $n\eta_K \to \infty$, the estimator $\boldsymbol{\hat \psi}$ is asymptotically normal with $\sqrt{K}(\boldsymbol{\hat \psi} - \boldsymbol{\psi}^*) \to \cN(0,\Sigma)$. A consistent estimator of $\Sigma$ is given by
$$
\hat\Sigma = \frac1{K} \sum_{k=1}^K (\boldsymbol{\hat\psi}_k-\boldsymbol{\hat\psi})^\top (\boldsymbol{\hat\psi}_k-\boldsymbol{\hat\psi}).
$$

\end{prp}

\begin{rem} $\hat{\boldsymbol{\psi}}$ is the same estimator as in Section~\ref{sec:pwpop} with $\alpha_k=1/K$ or in Section~\ref{sec:pwcenter} with $\beta_k=1/K$. However, they do not target the same estimand. In particular the asymptotic variance of the current $\hat{\boldsymbol{\psi}}$ is of order $1/K$ because of the between-study variance, while the estimators of Sections~\ref{sec:pwpop}-\ref{sec:pwcenter} only suffer from the within-study variance of order $1/(n\eta_K K) = o(1/K)$ under the assumption that $n\eta_K \to 0$.
\end{rem}

Table~\ref{tab:estimators} summarizes the four settings considered in this section. In the fully homogeneous setting---homogeneous populations with no center effect---the resulting estimator amounts to pooling observations across centers followed by a simple average. Under heterogeneous populations with no center effect, one first computes within-center averages and then combines them through a weighted average that can reflect, for instance, the relative sizes of subgroups. When populations are homogeneous but a center effect is present, the estimator takes a weighted average of per-center means; one may then choose, for example, to assign larger weights to more reliable centers. Finally, the fully heterogeneous setting is addressed via a hierarchical model; under the assumption that recruitment is uncorrelated with center effects, this structure allows population-level heterogeneity and center-level effects to be disentangled, leading to an estimator that uniformly averages the per-center means.

\begin{table}[h!]
\centering
\begin{tblr}{
  colspec = {X[1.35,l,m] X[1,c,m] X[1,c,m]},
  rowsep = 7pt,
  colsep = 8pt,
  vline{1} = {2-3}{0.8pt},
  vline{2,3} = {1-3}{0.8pt},
  vline{Z} = {1-3}{0.8pt},
  hline{1} = {2-3}{0.8pt},
  hline{2,3} = {1-3}{0.8pt},
  hline{Z} = {1-3}{0.8pt},
}
 & \textbf{No center effect} & \textbf{Center effect} \\
\textbf{Homogeneous population}
&
\(\displaystyle \hat{\psi}^a := \frac{n^{a1}}{n^a}\)
&
\(\displaystyle \hat{\psi}^a := \sum_k \beta_k \frac{n_k^{a1}}{n_k^a}\)
\\
\textbf{Heterogeneous population}
&
\(\displaystyle \hat{\psi}^a := \sum_k \alpha_k \frac{n_k^{a1}}{n_k^a}\)
&
\(\displaystyle \hat{\psi}^a := \frac{1}{K}\sum_k \frac{n_k^{a1}}{n_k^a}\)
\\
\end{tblr}
\caption{Estimators by population and center effect assumptions.}
\label{tab:estimators}
\end{table}

\section{Causal network meta-analysis} \label{sec:causalnma}

We now extend the causal framework of Section~\ref{sec:causalpair} to the network meta-analysis setting. Let $\cA=\{0,\dots,N\}$ denote the set of treatments and let $H\in[K]$ denote the study indicator. As in Section~\ref{sec:Reminders}, each study $k$ includes only a subset of treatments, denoted by  
$$\cA_k := \lbrace a\in \mathcal{A}: n_k^a \neq 0\rbrace, 
$$and, for each treatment $a\in\cA$, we denote by 
$$ \cK^a := \lbrace k \in [K]: n_k^a \neq 0\rbrace,
$$
the set of studies in which treatment $a$ is observed. Like in the pairwise setting, we consider this time a collection $(Y^a)_{a\in[N]}$ of counterfactuals and introduce once more the outcome functions
$$
\mu_k(a,x) := \bbE[Y^a \mid H=k,X=x], \quad \forall a \in [N], \forall x \in \cX, 
$$
so that the contrast reported in study $k$ between two treatments $a,b \in \cA_k$ is of the form
$$
\theta_k^{ab} := \Phi(\bbE_{P_k}[\mu_k(a,X)],\bbE_{P_k}[\mu_k(b,X)]) := \theta^{ab}(\mu_k,P_k).  
$$
We let again
$$
\psi_k^a := \bbE_{P_k}[\mu_k(a,X)], 
$$
be the study-specific absolute treatment effect of treatment $a$, and we aim at estimating contrasts of the form 
$$
\theta^{ab} = \theta^{ab}(\mu^*,P^*),
$$
for some specific target population $P^*$ and specific outcome function $\mu^*$.

\subsection{Causal NMA with homogeneous study populations and no center-effect}

We first consider the idealized setting in which all study populations are the same and share the same response function. Namely, we assume that 
$$P_k=P^* \quad\text{and}\quad \mu_k=\mu^* \quad \text{for all}\quad k\in[K]. 
$$
Under this assumption, it holds
$$ \psi_k^a = \bbE_{P^*}[\mu^\star(a,X)] =:\psi^{*a} $$
for every study $k$.

The only condition required in this setting is that each treatment is observed with positive probability:  
\begin{ass}[Treatment positivity for NMA] \label{ass:postrt_nma} For all $a\in \cA$, $\bbP(A=a)>0$.
\end{ass}

This condition is equivalent to requiring that $n^a:=\sum_{k\in\mathcal{K}_a}n_k^a$ is asymptotically nonzero for every treatment of interest. Under the latter assumption, the absolute causal effect of treatment $a$ in the common target population is identified by 
$$\psi^{*a} = \mathbb{E}[Y^a] = \mathbb{E}[Y\mid A=a],
$$
and is naturally estimated by the pooled arm-level estimator $$\hat\psi^a := \frac{n^{a1}}{n^a} = \frac{\sum_{k\in\mathcal{K}_a} n_k^{a1}} {\sum_{k\in\mathcal{K}_a} n_k^a}.$$ 

\begin{rem}
   This estimator coincides with the maximum likelihood estimator derived from a fixed-effect arm-based model when the link function is taken to be the identity, $h = \text{id}$ (see \eqref{eq:FE_MLE}). 
\end{rem}

The causal contrast between treatments $a$ and $b$ is then estimated by 
$$
\hat\theta^{ab} := \Phi(\hat\psi^a,\hat\psi^b). $$ 
In this setting, no direct or indirect comparison between $a$ and $b$ is required for $\theta^{ab}$ to be estimated. If treatments $a$ and $b$ are each observed somewhere in the network, then both $\psi^{*a}$ and $\psi^{*b}$ are identifiable and the contrast $\Phi(\psi^{*a},\psi^{*b})$ can be estimated, even in the absence of a path between $a$ and $b$ in the treatment network.

\begin{prp} \label{prp:variance_homo} Under Assumption~\ref{ass:sutva}-\ref{ass:nocenter} and \ref{ass:postrt_nma}, as $n \to \infty$, $\boldsymbol{\hat\psi}$ is asymptotically normal with $\sqrt{n}(\boldsymbol{\hat\psi}-\boldsymbol{\psi^*}) \to \cN(0,\Sigma)$ where $\Sigma$
is a diagonal matrix. A consistent estimator of $\Sigma$ is given by 
$$
\hat\Sigma^{aa} = \frac{n\hat\psi^a(1-\hat\psi^a)}{n^a}. 
$$
\end{prp}
The proof is straightforward and follow, for instance, similar lines as the proof of Proposition~\ref{prp:variance_pop} found in Appendix~\ref{app:variance_pop}.

\subsection{Causal NMA with heterogenous study populations or with center-effect}\label{sec:nma_heteroge&center_eff}

We now consider the more realistic setting in which both the study populations and the response functions may vary across studies. As in Section~\ref{sec:pwhetero}, we model this heterogeneity hierarchically.

\paragraph{Hierarchical model for NMA.} There is a law $\Pi$ on triplets $(P,\mu,e)$, where  \begin{itemize}[noitemsep]
\item $P\in\mathcal{P}(X)$ is a covariate distribution;
\item $\mu: \cA \times \cX \to [0,1]$ is the response function;
\item $e = (e^a)_{a\in \cA} \in \cP(\cA)$ is the assignment probability, parametrizing the probability of 
$A=a\mid H$.
\end{itemize}
Independently across studies $k=1,\dots,K$,
\[
(P_k,\mu_k,e_k)\ \sim\ \Pi.
\]
Given $(P_k,\mu_k,e_k)$, a unit $(X,A,Y)$ in study $k$ is generated by
\begin{equation}\label{eq:distrbs_givenH}
    X\mid H=k \sim P_k,\qquad
\bbP(A=a\mid H=k) = e_k^a,\qquad
Y^a \mid X=x,H=k \sim \mathrm{Ber}(\mu_k(a,x)),
\end{equation}
and, similar to the previous sections, we assume that Assumption~\ref{ass:sutva} holds, that is $Y=\sum_{a=0}^N \mathbf{1}(A = a)\, Y^a$. In this case, the positivity requirement is that every treatment has positive probability of appearing in each study in the network: \begin{ass}[Treatment positivity for NMA II]\label{ass:nma_arm_pos} There exists $\ve > 0$ such that for every $a\in\mathcal{A}$, $\Pi(e^a \geq \ve)>0$ and $\Pi(0 < e^a < \ve) = 0$. \end{ass}
This assumption ensures that $|\mathcal{K}_a|\to\infty$ as $K\to\infty$ for every treatment $a$. As in the pairwise case with both population heterogeneity and center effects, we define the mean response function and mean target population by 
$$\mu^\star(a,x):=\mathbb{E}_\Pi[\mu_k(a,x)], \qquad P^\star(B):=\mathbb{E}_\Pi[P_k(B)]
$$
for every measurable set $B$. The corresponding target absolute effect is 
$$\psi^{*a} := \mathbb{E}_{P^\star}[\mu^\star(a,X)].
$$
To identify this quantity from the studies in which treatment $a$ is actually observed, arm availability must not be informative about the latent population or response mechanism. This yields the following assumption.

\begin{ass}\label{ass:nma_mcar} It holds under $\Pi$ and for all $a \in \cA$, $\ind\{e^a>0\}$, $P$ and $\mu$ are mutually independent.
\end{ass}

The first part of Assumption~\ref{ass:nma_mcar} states that the inclusion of treatment $a$ in a study is unrelated to the covariate distribution and unrelated to the response function of that study. The second part is the same independence condition used in Section~\ref{sec:pwhetero}: center effects and population heterogeneity are allowed, but the recruitment and response mechanisms are independent from each other. Under Assumption~\ref{ass:nma_mcar}, for every treatment $a \in \cA$,  \begin{align*} \mathbb{E}_\Pi[\psi_k^a\mid e_k^a>0] &= \mathbb{E}_\Pi\left[ \mathbb{E}_{P_k}[\mu_k(a,X) \mid e_k^a > 0] \right] = \mathbb{E}_\Pi\left[ \mathbb{E}_{P_k}[\mu_k(a,X)] \right]=\mathbb{E}_{P^\star}[\mu^\star(a,X)] \\
&= \psi^{*a}. 
\end{align*}
This leads to the estimator 
$$
\hat\psi^a := \frac{1}{|\mathcal{K}_a|} \sum_{k\in\mathcal{K}_a} \hat\psi_k^a = \frac{1}{|\mathcal{K}_a|} \sum_{k\in\mathcal{K}_a} \frac{n_k^{a1}}{n_k^a},
$$
and, for any pair $a,b\in\mathcal{A}$, $$
\hat\theta^{ab} := \Phi(\hat\psi^a,\hat\psi^b). $$
Consistency follows easily from Proposition~\ref{prp:consistency_hier}. Regarding the variance, we find:
\begin{prp} \label{prp:variance_hetero} In the hierarchical model for NMA with Assumptions~\ref{ass:nma_arm_pos} and \ref{ass:nma_mcar},
in the regime where $K,n \to \infty$ with $\eta_K n \to \infty$, the vector $\hat{\boldsymbol{\psi}} = (\hat \psi^a)_{a \in \cA}$ is asymptotically normal with $\sqrt{K}(\hat{\boldsymbol{\psi}}-\boldsymbol{\psi^*}) \to \cN(0,\Sigma)$. A consistent  estimator for $\Sigma$ is given by
$$
\widehat \Sigma := \frac{1}{K} \sum_{k=1}^K [R_k\otimes (\boldsymbol{\hat\psi}_k-\boldsymbol{\hat\psi})]^\top [R_k \otimes (\boldsymbol{\hat\psi}_k-\boldsymbol{\hat\psi})],
$$
where 
$$
R_k^a := \frac{K}{|\cK^a|} \ind\{k \in \cK^a\},
$$
and where $\otimes$ denotes entry-wise multiplication of vectors.
\end{prp}

As an easy corollary, we find:

 \begin{cor}
 For a link-based contrasts $\Phi(x,y)=h(x)-h(y)$ with $h$ differentiable, an estimator of the variance of $\hat\theta^{ab}$ is given by
 $$ 
 (\hat\sigma^{ab})^2 :=   h'(\hat\psi^a)^2 \hat \Sigma^{aa} +  h'(\hat\psi^b)^2 \hat \Sigma^{bb} - 2  h'(\hat\psi^a)h'(\hat\psi^b) \hat \Sigma^{ab}.
 $$
 \end{cor}
The proof is a simple application of the $\Delta$-method.
\paragraph{Link with missing-data.} NMA can be formally as a missing-data problem. For each treatment $a$, the absolute effect $\psi_k^a$ is observed only when $M_k^a=1$ where $M_k^a := \ind\{e_k^a > 0\}$.  Assumption~\ref{ass:nma_mcar} corresponds to a missing completely at random condition: whether treatment $a$ appears in a study is independent of the latent population and response function of that study. Under this condition, the unweighted arm-level average over $\mathcal{K}_a$ estimates the common-target effect $\psi^{*a}$. A weaker missing at random condition would allow treatment availability to depend on observed study-level covariates $W_k$, such as publication year, country, disease severity criteria, or risk-of-bias indicators, but not on the unobserved components of $(P_k,\mu_k)$ after conditioning on $W_k$. In that case, a weighted estimator of the form $$
 \hat\psi_{\mathrm{MAR}}^a =  \sum_{k\in\mathcal{K}_a} \omega_k^a \hat\psi_k^a \quad \text{where}\quad \omega_k^a \propto \frac{1}{\mathbb{P}(M_k^a=1\mid W_k)}, 
 $$
 could target a common population, provided that the covariates $W_k$ are sufficiently informative.  Finally, if treatment availability depends on unobserved effect modifiers, unobserved baseline risks, or unobserved center-specific response mechanisms, then the missingness is informative. In this case, the estimator \[ \hat\psi^a = \frac{1}{|\mathcal{K}_a|} \sum_{k\in\mathcal{K}_a} \hat\psi_k^a \] is still a consistent estimator of $\bbE_\Pi[\psi_k^a \mid M_k^a =1]$ but the latter doesn't target a meaningful causal quantity anymore, and the contrast \[ \Phi\left( \mathbb{E}_\Pi[\psi_k^a\mid M_k^a=1], \mathbb{E}_\Pi[\psi_k^b\mid M_k^b=1] \right) \] then generally compares treatments across different target populations or response mechanisms.

\paragraph{Link with collapsibility.} The classical CL estimator of Section~\ref{sec:Reminders} is linear,
\begin{equation*}\label{eq:cl_linear}
    \hat{\mathbf d} = M\hat{\boldsymbol\theta}, \qquad M := \bigl(Z^\top W Z\bigr)^{+} Z^\top W,
\end{equation*}
with $W = \Sigma^{-1}$ (FE) or $W = (\Sigma + \hat\tau^2 V)^{-1}$ (RE), so each pooled contrast is a convex combination $\sum_k m_k\,\theta_k^{ab}$ of the study-specific ones. Whether this pooled quantity retains a causal meaning depends on the \emph{collapsibility} of the effect measure --- whether a population-level effect can be recovered from within-strata (here, per-study) effects.
\begin{itemize}[noitemsep,topsep=0pt]
    \item The risk difference is \emph{directly collapsible}:
\begin{equation*}\label{eq:rd_collapsible}
\theta_{\mathrm{RD}} = \bbE[Y^1] - \bbE[Y^0] = \bbE\big[\theta_{\mathrm{RD}}(H)\big].
\end{equation*}
CL pooling matches this: for convex weights, $\sum_k m_k\,\theta_{\mathrm{RD}}(k) = \bbE_{P^\star}[Y^1-Y^0]$ with $P^\star := \sum_k m_k\, P_k$ is again a risk difference. Any convex $m_k$ keeps a causal reading; only the target population changes.
\item The risk ratio is collapsible but not \emph{directly} so, requiring a weighted average:
\begin{equation*}\label{eq:rr_collapsible}
\theta_{\mathrm{RR}} = \frac{\bbE[Y^1]}{\bbE[Y^0]} = \bbE\!\left[\theta_{\mathrm{RR}}(H)\, w(H)\right], \quad w(k) := \frac{\bbE[Y^0\mid H = k]}{\bbE[Y^0]}.
\end{equation*}
The CL pooling still keeps a causal reading, now for $P^\star = \sum_k \alpha_k P_k$ with $\alpha_k = m_k/w(k)$. Unlike the RD case, this target population is identifiable only when the $w(k)$ are known, i.e. when the baseline risk $\bbE[Y^0\mid H=k]$ can be recovered --- which holds if the reference arm $a=0$ appears in every study.
    \item The odds ratio is non-collapsible: no weighting of conditional odds ratios reproduces the marginal one \citep{colnet2023risk}. The log scale changes nothing, as log RR and log OR remain non directly-collapsible.
\end{itemize}
\section{Numerical experiments} \label{sec:simulations}

The aims of this section are twofold. First, synthetic experiments illustrate the paper's main message. Classical fixed- and random-effects (network) meta-analytic procedures are consistent for the RD, but for non-linear measures these summaries cannot be interpreted as causal contrasts on an explicit target population. Second, we assess the finite-sample behavior of the proposed arm-level estimators and of the between-study variance estimators of Section~\ref{sec:nma_heteroge&center_eff}. All experiments are implemented in \textsf{R}: classical pairwise models are fitted with \texttt{metafor} (REML for $\tau^2$, \citealp{viechtbauer2005bias}) and classical network models with \texttt{netmeta} \citep{rucker2012network}; the causal estimators require only a few lines of code. The full simulation code is provided as supplementary material.

\subsection{Simulation design} \label{sub:simu_design}

\paragraph{A hierarchical data-generating process.} We simulate from an explicit instance of the hierarchical model introduced in Section~\ref{sec:pwhetero} and reused in Section~\ref{sec:nma_heteroge&center_eff}. The design associates one interpretable parameter with each source of heterogeneity, so that each of the four settings of Table~\ref{tab:estimators} is recovered by switching parameters on or off. Covariates take values in $\bbR^d$ with $d=2$ throughout. Independently across studies $k \in [K]$, we draw
\begin{equation}\label{eq:dgp_study_level}
m_k \sim \cN\big(0,\tau_{\mathrm{pop}}^2 I_d\big), \qquad
u_k \sim \cN(0,\sigma_u^2), \qquad
v_k^a \overset{\text{iid}}{\sim} \cN(0,\sigma_v^2), \qquad
e_k = \varepsilon\,\mathbf{1} + \big(1-(N{+}1)\varepsilon\big)\,D_k,
\end{equation}
where $D_k \sim \mathrm{Dirichlet}(\kappa,\dots,\kappa)$ and all draws are mutually independent. The triple $(P_k,\mu_k,e_k) \sim \Pi$ of the hierarchical model is then given by the covariate distribution $P_k := \cN(m_k,\sigma_X^2 I_d)$, the assignment probabilities $e_k$, and the response function
\begin{equation}\label{eq:dgp_response}
\mu_k(a,x) := \mathrm{expit}\big(\gamma_a + \lambda_a^\top x + u_k + v_k^a\big),
\qquad \mathrm{expit}(z) = (1+e^{-z})^{-1},
\end{equation}
where the parameters $(\gamma_a,\lambda_a)_{a \in \cA}$ are shared across studies, drawn once and for all ($\gamma_a \sim \cN(0,s_\gamma^2)$ and $\lambda_a$ with i.i.d.\ $\cN(0,s_\lambda^2)$ entries) and held fixed across Monte-Carlo replications, so that all replications share a common outcome model. Patient-level data are then generated i.i.d.\ as in Section~\ref{sec:pwhetero}: $H \sim \mathrm{Unif}([K])$ (so that $\bbP(H=k)=1/K$ and $\eta_K = 1/K$), $X \mid H=k \sim P_k$, $A \mid H=k \sim \mathrm{Categorical}(e_k)$, $Y^a \mid X=x, H=k \sim \mathrm{Bernoulli}(\mu_k(a,x))$ independently across arms, and $Y = Y^A$, so that Assumptions~\ref{ass:sutva} and~\ref{ass:rcts} hold by construction. Only the arm-level counts $n_k^{ay}$ --- that is, the data of Tables~\ref{tab:rct2} and~\ref{tab:nma} --- are passed to the estimators; individual covariates, assignment probabilities and counterfactuals are discarded.

\paragraph{Mapping between simulation parameters and assumptions.} Each parameter of \eqref{eq:dgp_study_level}--\eqref{eq:dgp_response} activates one and only one structural feature of the causal framework:
\begin{itemize}[noitemsep]
    \item $\tau_{\mathrm{pop}}$ tunes \emph{population heterogeneity} (the arrow $H \to X$): $\tau_{\mathrm{pop}}=0$ makes all $P_k$ equal, i.e., Assumption~\ref{ass:homop} holds. One can note that because the slopes $\lambda_a$ differ across arms, population differences  translate into genuine treatment-effect modification across studies, not mere baseline shifts.
    \item $(\sigma_u,\sigma_v)$ tune the \emph{center effect} (the arrow $H \to Y$): $u_k$ shifts the outcome level of study $k$ uniformly across arms (e.g., differences in standards of care or outcome measurement), while $v_k^a$ is a study-by-treatment interaction (treatment $a$ delivered slightly differently at center $k$). Setting $\sigma_u=\sigma_v=0$ recovers the no-center-effect Assumption~\ref{ass:nocenter}.
    \item The draws $m_k$ and $(u_k,v_k)$ are independent, hence $P_k \indep \mu_k$: Assumption~\ref{ass:response_cov_indep} -- recruitment carries no information about the outcome mechanism -- holds by construction 
    \item The floor $\varepsilon$ guarantees $e_k^a \geq \varepsilon$ almost surely, so uniform treatment positivity holds; the concentration $\kappa$ produces moderately unbalanced allocations across studies, in line with Assumption~\ref{ass:pos2}.
\end{itemize}
In all experiments we set $\sigma_X = 1$, $s_\gamma = 0.7$, $s_\lambda = 0.8$, $\varepsilon = 0.05$ and $\kappa = 5$.

\paragraph{From complete networks to NMA data.} For the network experiments, the generator first produces a complete network in which every study carries all $N+1$ arms; a masking step then hides arms so as to reproduce the incomplete designs of Section~\ref{sec:causalNMA}. One anchor treatment ($a=0$) is kept in every study, and every other arm is retained independently with probability $p$, the retention indicators being drawn independently of $(P_k,\mu_k,e_k)$. This construction has three consequences: (i) every treatment is reported by a positive fraction of studies for a big enough $K$, so Assumption~\ref{ass:nma_arm_pos} holds; (ii) arm availability is independent of the latent population and response mechanisms, which instantiates exactly the MCAR condition of Assumption~\ref{ass:nma_mcar}; (iii) the network is star-shaped and thus connected by construction (although, as emphasized in Section~\ref{sec:causalnma}, connectivity plays no role in the causal estimator). %Masking only hides counts that were actually generated; it never modifies the data.

\paragraph{Ground truth.} Since the simulator generates the full vector of potential outcomes $(Y^a)_{a\in\cA}$ for every unit, causal estimands can be read directly off the data. To evaluate a target absolute effect $\psi^{*a}$ we draw a large auxiliary population ($5\times 10^4$ units in the pairwise experiment, $5\times 10^5$ in the network experiments) while re-using the same latent draws $(P_k,\mu_k,e_k)$ as the observed dataset (same random seed), and average the simulated $Y^a$ over it. How the auxiliary population is drawn depends on the estimand, and the mixture settings and the hierarchical setting play different roles here.
\begin{itemize}[noitemsep]
    \item In the mixture settings of Sections~\ref{sec:pwpop} (heterogeneous populations, no center effect) and~\ref{sec:pwcenter} (center effect, homogeneous population), the weights are a genuine modeling choice. One \emph{specifies} the target population $P^\star = \sum_k \alpha_k P_k$ (resp.\ the target response $\mu^\star = \sum_k \beta_k \mu_k$) and samples $H \sim \mathrm{Categorical}(\alpha)$ (resp.\ $\beta$), so that the truth is evaluated on the chosen mixture $\psi^{*a} = \sum_k \alpha_k \psi_k^a$.
    \item In the hierarchical setting of Section~\ref{sec:pwhetero} (heterogeneous populations \emph{and} a center effect), the weights are not chosen. The target is the population-level mean $\bbE_\Pi[\psi_k^a] = \bbE_{P^\star}[\mu^\star(a,X)]$ of Equation~\ref{eq:mean_baseline}, and the estimator is $\hat\psi^a = K^{-1}\sum_k \hat\psi_k^a$. The matching sampling scheme is therefore the uniform one, $H \sim \mathrm{Unif}([K])$, and $w_k = 1/K$ is imposed by the model rather than selected.
\end{itemize}
The pairwise scenario below is of this hierarchical type; the two network scenarios also use $w_k = 1/K$. Because the seed is shared, the auxiliary population carries the same $(P_k,\mu_k)$ as the data, so the red truth line is exactly the estimand attached to the $K$ studies at hand.
\paragraph{Estimators under comparison.} In the pairwise experiment we compare (FE) and (RE) inverse-variance pooling of the per-study contrasts as in \eqref{eq:ivw}, with $\tau^2$ estimated by REML, against the causal estimator of Section~\ref{sec:pwhetero}, $\hat\theta = \Phi(\hat\psi^1,\hat\psi^0)$ with $\hat\psi^a = K^{-1}\sum_k n_k^{a1}/n_k^a$. In the network experiments we compare the contrast-likelihood models CL-FE and CL-RE of Section~\ref{sec:Reminders} (the generalized least-squares solutions $\hat{\mathbf d}^{\rm FE}$ and $\hat{\mathbf d}^{\rm RE}$, as implemented in \texttt{netmeta}) against the causal estimator of Section~\ref{sec:nma_heteroge&center_eff}, $\hat\psi^a = |\cK^a|^{-1}\sum_{k\in\cK^a} n_k^{a1}/n_k^a$, contrasts again being formed after averaging. Confidence intervals for the causal contrasts are the Wald intervals obtained from the between-study covariance estimator $\hat\Sigma$ and the delta-method variance $(\hat\sigma^{ab})^2$ derived at the end of Section~\ref{sec:nma_heteroge&center_eff}; a continuity correction of $1/2$ is applied to per-study rates to guard against zero cells (\texttt{netmeta} applies its own correction). All network contrasts are reported against the anchor. Table~\ref{tab:simu_scenarios} summarizes the three scenarios presented below; note that they occupy, respectively, the bottom-right cell of Table~\ref{tab:estimators}, the idealized setting of Section~\ref{sec:causalnma}, and its population-heterogeneous extension.

\begin{table}[h!]
\centering
\begin{tabular}{lccccccc}
\toprule
Scenario & $\tau_{\mathrm{pop}}$ & $\sigma_u = \sigma_v$ & $K$ & $N{+}1$ & $n$ & Masking & Repl. \\
\midrule
Pairwise, het.\ pop.\ $+$ center effect & $1$ & $0.5$ & $10$ & $2$ & $4\,000$ & --- & $200$ \\
NMA, homogeneous, no center effect & $0$ & $0$ & $10$ & $5$ & $6\,000$ & MCAR, $p=0.6$ & $200$ \\ % CHECK: K, N+1, n
NMA, het.\ populations, no center effect & $1$ & $0$ & $10$ & $5$ & $6\,000$ & MCAR, $p=0.6$ & $200$ \\ % CHECK: K, N+1, n
\bottomrule
\end{tabular}
\caption{Simulation scenarios. In all cases $d=2$, $\sigma_X=1$, $s_\gamma=0.7$, $s_\lambda=0.8$, $\varepsilon=0.05$, $\kappa=5$, uniform weights $w_k = 1/K$ and uniform study membership, so that the expected study size is $n/K$. ``Repl.'' is the number of Monte-Carlo replications; all scenarios, pairwise and network alike, are evaluated over these replications, with the truth recomputed each replication on the matched latent draws.}
\label{tab:simu_scenarios}
\end{table}

\subsection{Pairwise meta-analysis with population heterogeneity and center effects} \label{sub:simu_pairwise}

\begin{figure}[h!]
    \centering
    \includegraphics[width=\textwidth]{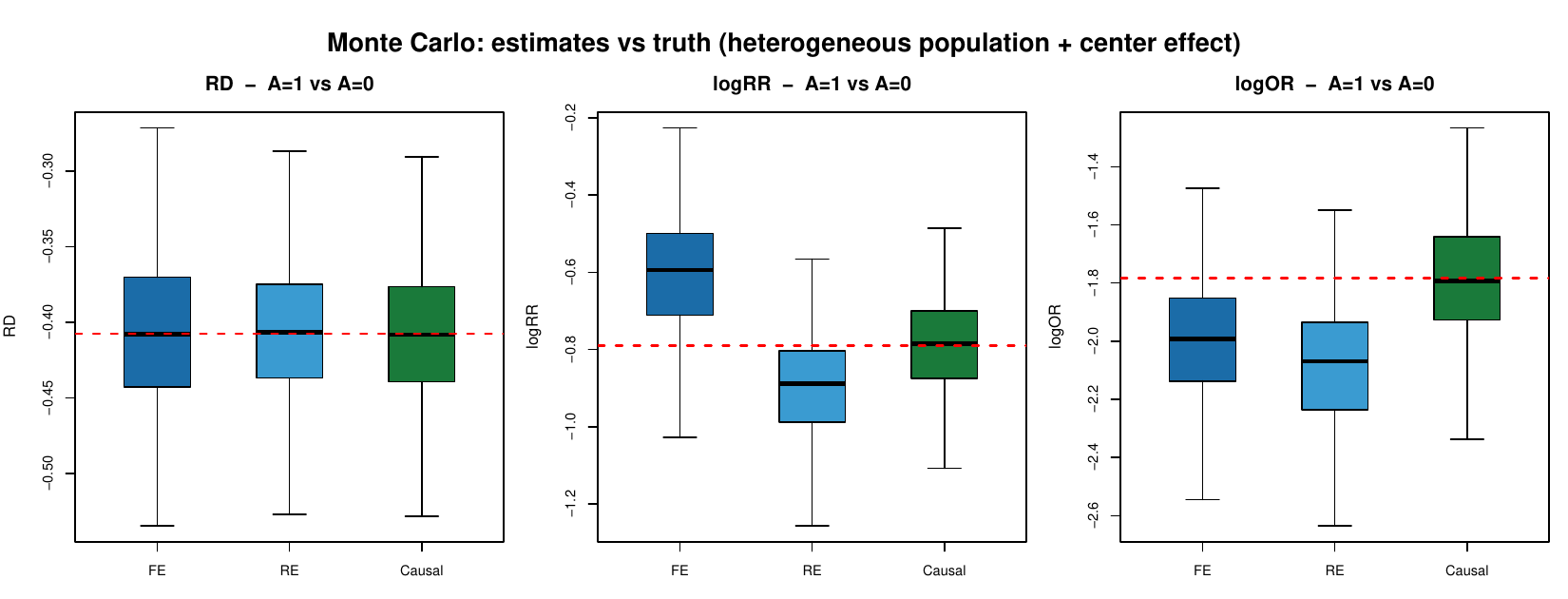}
    \caption{Monte-Carlo distributions ($B=200$ replications) of the FE, RE and causal estimators of the contrast between $A=1$ and $A=0$ in the fully heterogeneous pairwise setting ($\tau_{\mathrm{pop}}=1$, $\sigma_u=\sigma_v=0.5$, $K=10$, $n=4\,000$, uniform weights). Left: risk difference; middle: log risk-ratio; right: log odds-ratio. The red dashed line is the causal estimand $\theta(\mu^\star,P^\star)$ of Section~\ref{sec:pwhetero}, averaged over replications.}
    \label{fig:mc_pairwise_both}
\end{figure}

We first instantiate the fully heterogeneous pairwise setting of Section~\ref{sec:pwhetero}: study populations differ ($\tau_{\mathrm{pop}}=1$) \emph{and} a center effect is present ($\sigma_u=\sigma_v=0.5$), with $K=10$ studies of expected size $n\eta_K = 400$. Over $B = 200$ independent replications --- the latent draws $(P_k,\mu_k,e_k)$ being redrawn each time --- we compute the FE, RE and causal estimators of the contrast between $A=1$ and $A=0$ on the risk-difference, log risk-ratio and log odds-ratio scales. Figure~\ref{fig:mc_pairwise_both} displays the Monte-Carlo distributions; the red dashed line marks the causal estimand $\theta(\mu^\star,P^\star)$, averaged over replications.

Two features stand out. First, on the risk-difference scale the three estimators are essentially indistinguishable and all centered on the causal truth. This is expected: the risk difference is a linear --- hence directly collapsible --- contrast, for which averaging contrasts and contrasting averages coincide, so that classical pooling automatically inherits the causal interpretation; this is the pairwise phenomenon explained in \cite{berenfeld2025causal}, for whom the risk difference is the only standard measure whose classical aggregation is automatically causal. Second, on the two nonlinear scales the causal estimator remains centered on the truth while FE and RE are visibly biased --- and, importantly, not in a predictable direction: FE is attenuated toward the null on the log risk-ratio scale yet overshoots the truth on the log odds-ratio scale, while RE overshoots on both.
%\ab{a voir si on garde}
%Third, the dispersion of the causal estimator is comparable to that of RE: making the target population explicit costs nothing in precision here. (Part of the displayed spread of all three estimators reflects the replication-to-replication variability of the estimand itself, since the latent study draws are resampled in each replication.)
%\jj{En général, on dit plutôt qu'on a des plus petites variances} 

%\ab{c'est en pratique ça mais dans les simus je trouve pas trop}
\subsection{Causal NMA under homogeneity: a sanity check} \label{app:simu_nma_hom}

\begin{figure}[h!]
    \centering
    \includegraphics[width=\textwidth]{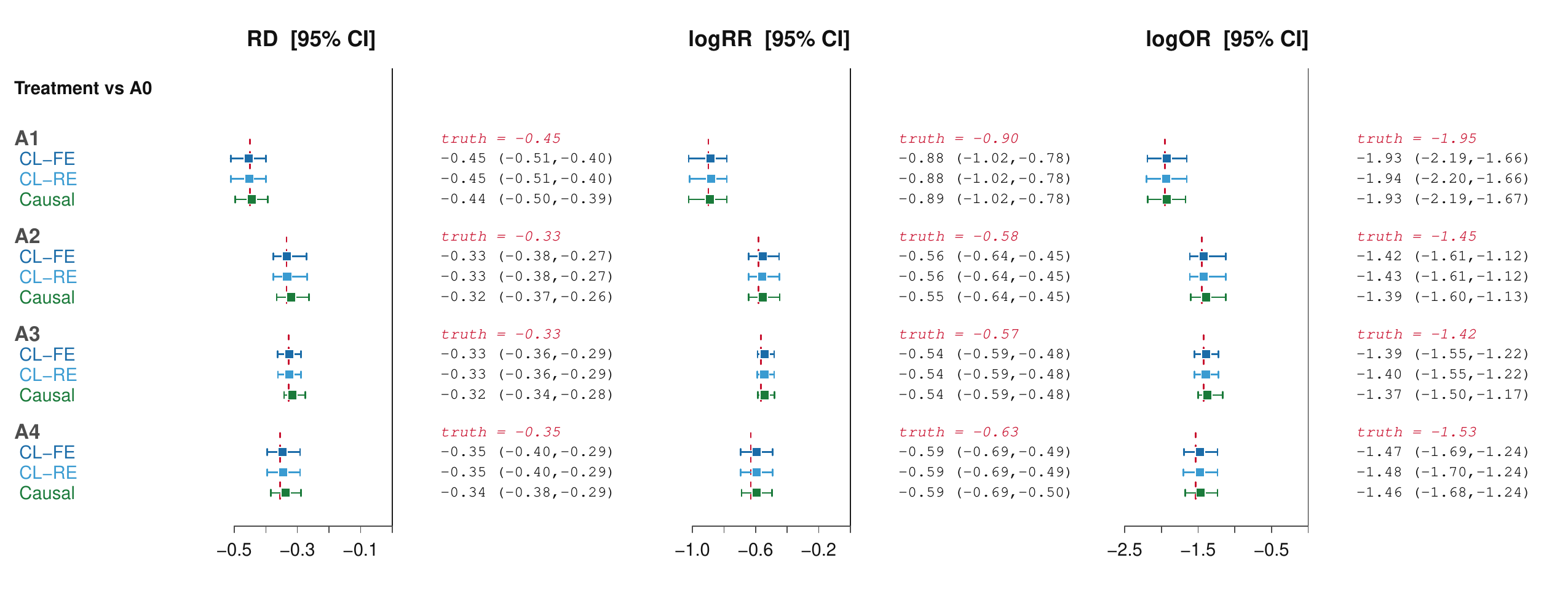}
   \caption{Network meta-analysis under homogeneous populations and no center effect ($\tau_{\mathrm{pop}}=0$, $\sigma_u=\sigma_v=0$; see Table~\ref{tab:simu_scenarios}), over $200$ Monte-Carlo replications. Each treatment is contrasted against the anchor $a=0$ on the RD, log risk-ratio and log odds-ratio scales; points are Monte-Carlo means and bars the $2.5$--$97.5\%$ range of the estimates, with the red dashed line the causal truth. CL-FE and CL-RE are fitted with \texttt{netmeta}; the causal estimator uses the between-study delta-method variance of Section~\ref{sec:nma_heteroge&center_eff}.}
    \label{fig:nma_hom}
\end{figure}

We next consider the idealized network setting of Section~\ref{sec:causalnma}, in which all studies share the same population and response function ($\tau_{\mathrm{pop}}=0$, $\sigma_u=\sigma_v=0$), with $N+1=5$ treatments, $K=10$ studies, $n=6\,000$ patients and MCAR masking with retention probability $p=0.6$ (anchor $a=0$ present everywhere). % CHECK: K, N+1, n
Figure~\ref{fig:nma_hom} reports, for a single masked dataset, the estimated risk-ratios of each treatment against the anchor, together with $95\%$ confidence intervals.

As anticipated, when the classical assumptions genuinely hold, all three procedures agree: there is a single vector of absolute effects $(\psi^{*a})_{a\in\cA}$, every estimator is consistent for the same contrasts, and the three point estimates coincide up to sampling noise with intervals of comparable width, all compatible with the truth. Two remarks are in order. First, the causal estimates were computed without ever forming the network graph: only the marginal availability of each arm ($|\cK^a| > 0$) was used, and under homogeneity the average-of-rates estimator of Section~\ref{sec:nma_heteroge&center_eff} and the pooled estimator $n^{a1}/n^a$ of Section~\ref{sec:causalnma} are both consistent for $\psi^{*a}$ and nearly equal numerically. Second, this scenario shows that adopting the causal formulation carries no cost in the ideal case: it simply reproduces the classical answer, while making explicit the population for which it is valid.

\subsection{Causal NMA  with population heterogeneity} \label{sub:simu_nma_het}
 We now focus on the network analogue of the population-mixture setting of Section~\ref{sec:pwpop}: study populations differ ($\tau_{\mathrm{pop}}=1$) but there is no center effect, with $N+1=5$ treatments, $K=10$ studies of expected size $n/K = 600$, and MCAR masking with $p=0.6$, so that each non-anchor arm is reported by roughly $0.6K \approx 6$ studies.

\begin{figure}[h!]
    \centering
    \includegraphics[width=\textwidth]{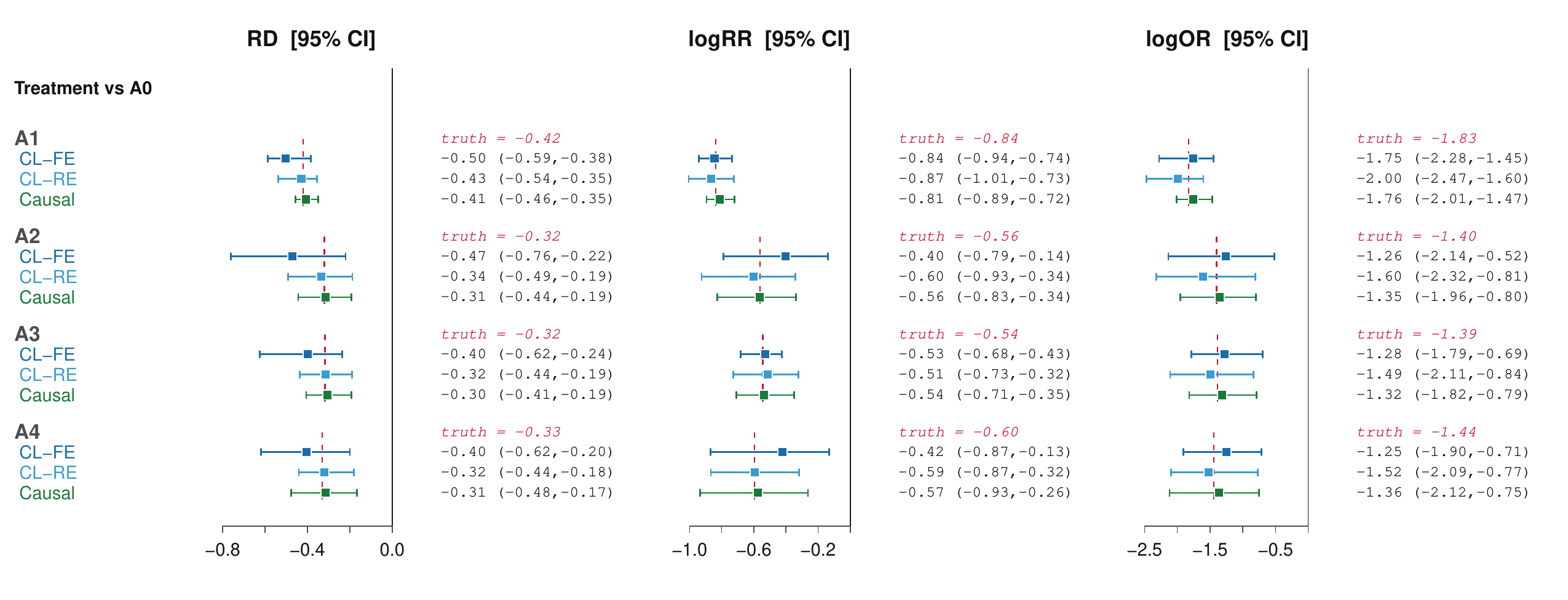}
    \caption{Network meta-analysis under heterogeneous populations and no center effect ($\tau_{\mathrm{pop}}=1$, $\sigma_u=\sigma_v=0$; see Table~\ref{tab:simu_scenarios}), over $200$ Monte-Carlo replications. Each treatment is contrasted against the anchor $a=0$ on the RD, log risk-ratio and log odds-ratio scales; points are Monte-Carlo means and bars the $2.5$--$97.5\%$ range, with the red dashed line the causal truth $\Phi(\psi^{*a},\psi^{*0})$ on the uniform mixture of the study populations.}
    \label{fig:nma_het}
\end{figure}

The picture is strikingly regular. First, the causal estimator is centered on the truth in all  contrasts, with intervals that cover it; its precision for the contrast $\theta^{ab}$ is governed by $|\cK^a|$ and $|\cK^b|$ --- the numbers of studies informing each arm --- rather than by the geometry of the network, and contrasts between two non-anchor treatments would be obtained in exactly the same way, whether or not the pair is ever compared head-to-head. 

Second, on the risk-difference scale the CL-RE and Causal are essentially indistinguishable and all centered on the causal truth, and the collapsibility discussion of Section~\ref{sec:nma_heteroge&center_eff} (Eqs.~\eqref{eq:rd_collapsible}--\eqref{eq:rr_collapsible}) makes precise why. Because the risk difference is \emph{directly} collapsible, any convex-weighted average of study-specific risk differences is itself a risk difference on the correspondingly weighted mixture of study populations; the classical estimand thus stays causal, and it coincides with the \emph{particular} truth plotted here --- the uniform mixture $w_k = 1/K$ --- exactly when the pooling weights are themselves uniform. This is what the design delivers: with $H\sim\mathrm{Unif}([K])$ the studies have equal expected sizes, so the RE inverse-variance weights converge toward the uniforme; both estimands therefore sit on the $1/K$ mixture and match the causal estimator. Had the pooling weights departed from $1/K$, RE would remain causal on this scale but would target a \emph{different} population and drift off the red line --- a target-population mismatch, not a loss of causal meaning. This is the risk-difference special case of the phenomenon of \cite{berenfeld2025causal}, for whom it is the only standard measure whose classical aggregation is automatically causal. No such rescue exists for the nonlinear measures: by Eq.~\eqref{eq:rr_collapsible} the pooling weights there must additionally absorb the baseline-risk collapsibility weights $w(X)$, which no study-level weighting can reproduce, so the bias in the middle and right panels persists regardless of the design. 

Finally, we note that because the causal approach delivers \emph{absolute} effects $(\hat\psi^a)_{a\in\cA}$ on a single explicit population, together with a joint asymptotic covariance, treatment rankings and ranking probabilities in the spirit of SUCRA \citep{salanti2011rankings} or p-scores \citep{Rucker2015Ranking} can be produced directly on the $\psi$ scale by simulating from the estimated Gaussian limit --- with the notable difference that the ranked quantities now carry an explicit causal meaning.

\begin{rem}[Studies, not patients, drive precision]
In every hierarchical scenario --- pairwise het.\ pop.\ $+$ center effect and all
network settings of Section~\ref{sec:nma_heteroge&center_eff} --- the CLT is in
$\sqrt{K}$, so $\hat\psi^a$ has variance $\Var_\Pi(\psi_k^a)/K$ (with
$|\cK^a|\approx pK$ in the network case): intervals shrink like $1/\sqrt{K}$ and
are insensitive to $n$ once $n\eta_K\to\infty$. Since $\Var_\Pi(\psi_k^a)$ grows
with both population heterogeneity ($\tau_{\mathrm{pop}}$) and the center effect
($\sigma_u,\sigma_v$), more heterogeneity means more \emph{studies} are needed
for a given precision.
\end{rem}

\begin{rem}[Estimand versus arm availability]
The causal estimator averages over the studies $\cK^a$ that report arm $a$, while the truth above is defined over the full mixture $\sum_k w_k (P_k,\mu_k)$. Under MCAR retention, the availability indicator $M_k^a$ is independent of $(P_k,\mu_k)$, so with uniform weights the available-study average is unbiased for the full-mixture estimand and the comparisons above are fair. Under an informative masking mechanism --- e.g., a retention probability depending on the population location $m_k$ --- the two quantities diverge: the estimator then consistently estimates $\bbE_\Pi[\psi_k^a \mid M_k^a = 1]$, which no longer corresponds to the intended target (see the missing-data discussion of Section~\ref{sec:nma_heteroge&center_eff}), and a bias appears against the truth line. Re-running the experiment under this mechanism thus provides a simple diagnostic template for sensitivity analyses to informative arm availability.
\end{rem}

\paragraph{Takeaways.} The three experiments give an empirical counterpart to the theory of Sections~\ref{sec:causalpair}--\ref{sec:causalnma}. When populations are homogeneous, or when the contrast is the (collapsible) risk difference, classical and causal analyses agree, and the causal formulation merely makes the target population explicit at no statistical cost. As soon as populations are heterogeneous and a nonlinear measure is used, the classical fixed- and random-effects summaries drift away from the causal contrast in directions that depend on the measure and on the design while the simple arm-level estimator, equipped with the between-study variance of Section~\ref{sec:nma_heteroge&center_eff}, remains centered with reliable uncertainty quantification, and does so without ever invoking the network graph or the transitivity assumption.

\section{Real-World Experiment}
\label{sec:realworld}

\begin{figure}[h!]
    \centering
    \includegraphics[width=\textwidth]{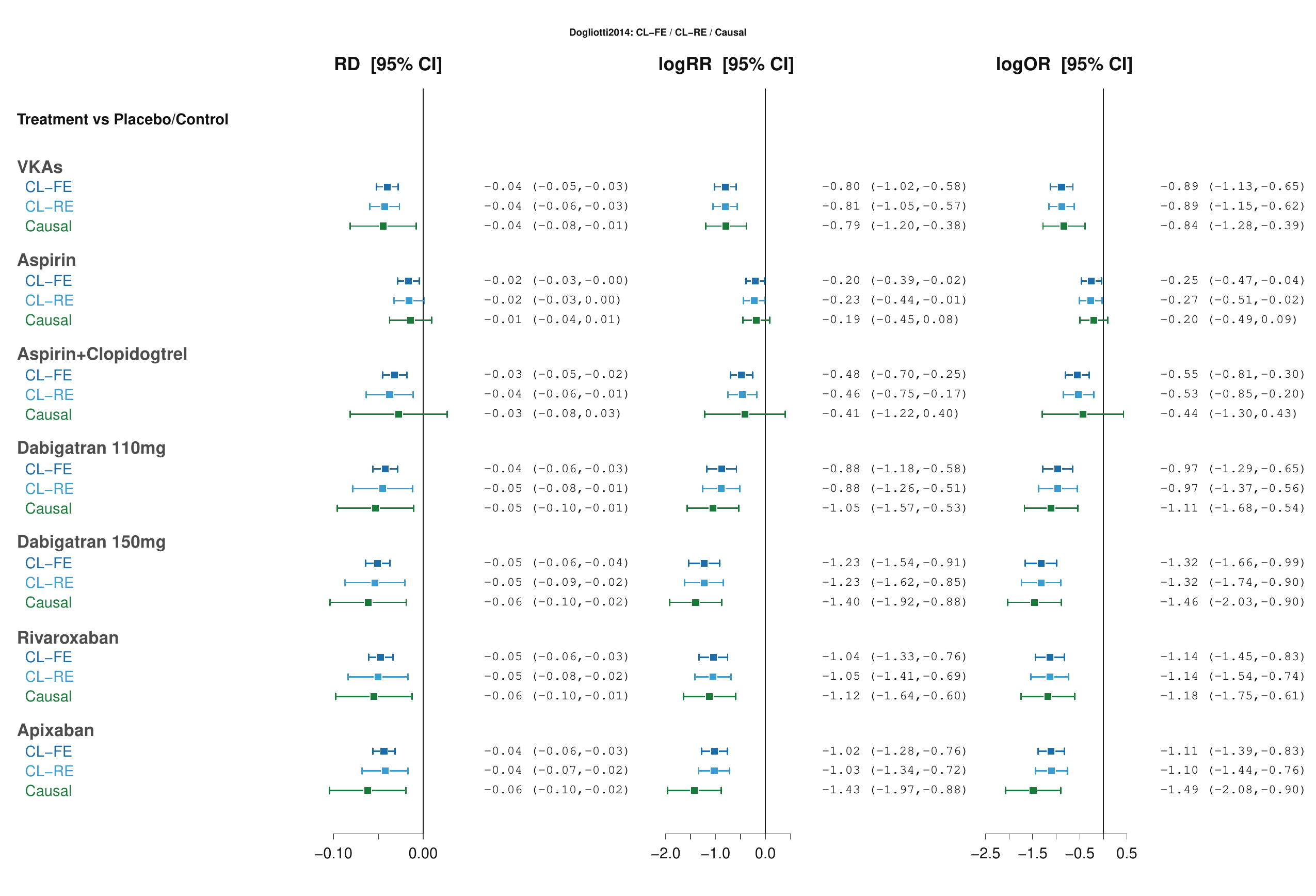}
    \caption{Re-analysis of the atrial-fibrillation network of \cite{dogliotti2014current}: each antithrombotic strategy is contrasted against control on the RD, log risk-ratio and log odds-ratio scales, using CL-FE, CL-RE (\texttt{netmeta}) and the causal arm-level estimator of Section~\ref{sec:nma_heteroge&center_eff}. Points are estimates and bars $95\%$ confidence intervals. The three methods agree on direction; the causal intervals are somewhat wider on this dense network.}
    \label{fig:rw_dogliotti}
\end{figure}

\begin{figure}[h!]
    \centering
    \includegraphics[width=\textwidth]{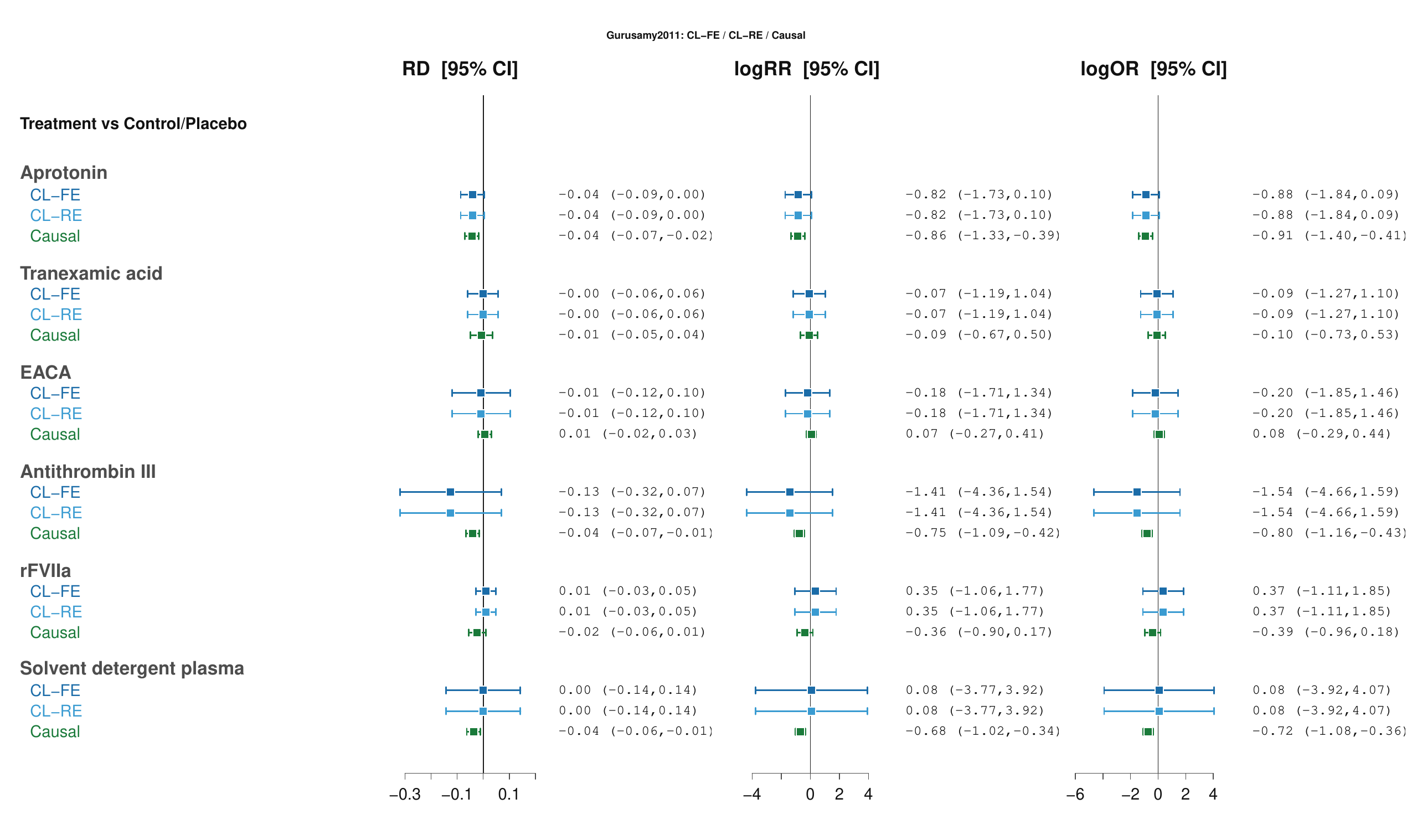}
    \caption{Re-analysis of the bleeding-reduction network of \cite{gurusamy2011methods}, same layout as Figure~\ref{fig:rw_dogliotti}. Here the network is sparse: CL-FE and CL-RE coincide (heterogeneity estimated at zero) and their intervals are very wide, while the causal intervals are markedly tighter and occasionally significant where CL is not.}
    \label{fig:rw_gurusamy}
\end{figure}

To complement the simulations, we re-analyse two published network meta-analyses using only the information our method needs, namely the arm-level outcome counts $n_k^{ay}$ of Table~\ref{tab:nma}. For each dataset we compare the contrast-likelihood estimators CL-FE and CL-RE (fitted with \texttt{netmeta}) against the causal arm-level estimator of Section~\ref{sec:nma_heteroge&center_eff}, on the risk-difference, log risk-ratio and log odds-ratio scales. Every treatment is contrasted against the common control/placebo arm, which plays the role of the reference; as in the simulations, the causal contrasts use the delta-method variance $(\hat\sigma^{ab})^2$, with a $1/2$ correction for zero cells.
The first dataset, from \cite{dogliotti2014current}, compares seven antithrombotic strategies for stroke prevention in atrial fibrillation (VKAs, aspirin, aspirin plus clopidogrel, dabigatran 110 and 150\,mg, rivaroxaban and apixaban) against placebo. The second, from \cite{gurusamy2011methods}, compares six interventions aimed at reducing perioperative bleeding (aprotinin, tranexamic acid, EACA, antithrombin III, rFVIIa and solvent-detergent plasma) against placebo. The two networks are of very different structure: the first is dense and based on large trials, the second is sparse and based on smaller studies. Figures~\ref{fig:rw_dogliotti} and~\ref{fig:rw_gurusamy} report the results.

Three points stand out. First, the three methods broadly agree on the direction and rough magnitude of every effect; on the risk-difference scale, where classical pooling can be causal, the point estimates are nearly identical across methods, exactly as in the simulations. Second, the methods differ mostly in the width of the confidence intervals, and not in a single direction. In the dense Dogliotti network the causal intervals are typically somewhat wider than the CL ones, occasionally enough to change a borderline conclusion, as for aspirin, whose log risk-ratio is significant under CL but has a causal interval covering zero ($-0.19$, CI $-0.45$ to $0.08$). In the sparse Gurusamy network the situation is reversed: the causal intervals are often much narrower. For antithrombin III the risk difference goes from a wide, non-significant CL interval ($-0.13$, CI $-0.32$ to $0.07$) to a tight, significant causal one ($-0.04$, CI $-0.07$ to $-0.01$); similar tightening occurs for solvent-detergent plasma and rFVIIa, where the very wide CL intervals even sit on the opposite side of zero from the causal point estimate.

The reason is simple and matches the theory: the precision of the causal estimator is driven by how many studies report each arm and by their sizes, not by the geometry of the network. When the network is rich, CL borrows strength across many indirect paths and can look tighter; when it is sparse, indirect comparisons inflate the CL variance while the plain arm-level average stays stable. One can also note that in the Gurusamy analysis CL-FE and CL-RE coincide, because the between-study heterogeneity is estimated at zero, so the random-effects model adds nothing there.

Since these are real data, the causal truth is unknown and we cannot know which method is the best. What the two examples do show is that the causal estimator behaves sensibly on both a dense and a sparse network, that it agrees with the classical answer whenever one would expect it to (risk difference, dense evidence), and that the discrepancies concentrate on the nonlinear scales and on the sparse networks--- again consistent with Sections~\ref{sec:causalpair}--\ref{sec:causalnma}. As always, the arm-level average targets an explicit average population under the MCAR-type condition of Assumption~\ref{ass:nma_mcar}, which is not testable from aggregate data and would warrant a sensitivity analysis using IPD.

\section{Conclusion}

Beyond introducing new estimators, the causal framework developed in this work provides a principled way to place clinically meaningful estimands at the center of meta-analysis and to clarify the interpretation of treatment effects.
However, this causal perspective  also highlights the intrinsic limitations of aggregate-data meta-analysis. Once heterogeneous trial populations and center effects are simultaneously acknowledged, the information available at the aggregate level inevitably restricts the range of identifiable causal estimands. 
In particular, the treatment effect can only be identified for a target population corresponding to a uniform average of the studies population, in a center that is the uniform average of the centers in the studies and there is no flexibility to  choose the weighting scheme. Nevertheless, the resulting estimator, which can be interpreted as an arm-level aggregation, is remarkably simple whatever the heterogeneity considered. In the network setting, however, it no longer coincides with any of the standard network meta-analysis estimators, despite the extensive methodological literature devoted to this problem.

The good properties of this estimator relies on the  assumption that treatment choices, center effects, and the underlying population composition are mutually independent. The plausibility of that assumption, as well as the impact of potential violations, should therefore be investigated through systematic sensitivity analyses. Whenever individual participant data are available, they provide a valuable opportunity to assess these assumptions empirically and to quantify the robustness of causal conclusions.

More broadly, this work suggests that aggregate data should be viewed as one point along a continuum of available information rather than as an endpoint. Incorporating richer sources of evidence data (such as, baseline characteristics reported in trial publications or site-level covariates) would allow less restrictive  assumptions. In particular, treatment assignment mechanisms depending on observed site characteristics could be easily accommodated within the same causal framework. 

Ultimately, a causal perspective suggests that the central question in meta-analysis is not how treatment contrasts or arms should be combined across studies, but rather which causal effect is to be estimated, in which target population, and under which identifying assumptions.

\paragraph{Acknowledgements.} 

We thank Anna Chaimani, Tim Morris, Antonio Remiro-Azócar, Erwan Scornet and Ian White.  for insightful exchanges that broadened our perspective on this topic.

This work has been done in the frame of the PEPR SN SMATCH project and has benefited from a governmental grant managed by the Agence Nationale de la Recherche under the France 2030 programme, reference ANR-22-PESN-0003.

\bibliography{references}

\appendix

\section{Proofs}

\subsection{Proof of Proposition~\ref{prp:variance_pop}} \label{app:variance_pop}
$\boldsymbol{\hat\psi}$ is asymptotically normal per the $\Delta$-method. Furthermore, conditional on the set of variables $\bn := \{n_k^a\}_{k\in[K],a\in\{0,1\}}$ and on the event $\cE := \{n_k^a > 0~,\forall k\in[K],a\in\{0,1\}\}$ (which probability goes to $1$ exponentially fast in $n$), the estimators $\hat\psi^1$ and $\hat\psi^0$ are independent from each other, and the $\hat\psi^a_k$'s are also mutually independent. Since $\hat\psi^a$ and the $\hat\psi^a_k$'s are centered conditionally on $\cE$ and $\bn$, we find that
\begin{align*}
n \Var \hat\psi^a &= n ~\bbE[\Var(\hat\psi^a\mid \cE,\bn)\mid \cE] + o(1) \\
&= n \sum_{k=1}^K \alpha_k^{*2} \bbE[\Var(\hat\psi^a_k\mid \cE,\bn)\mid \cE] + o(1) \\
&= n \sum_{k=1}^K \alpha_k^{*2} \psi_k^a(1-\psi_k^a)\bbE\left[\frac{1}{n_k^a}~\middle|~ \cE\right] + o(1) \\ 
&= \sum_{k=1}^K \alpha_k^{*2} \frac{\psi_k^a(1-\psi_k^a)}{\bbP(H=k,A=a)} + o(1),
\end{align*}
which ends the proof.

\subsection{Proof of Proposition~\ref{prp:consistency_hier}} \label{app:consistency_hier}

We reuse the notations ($\bn$,$\cE$) of the previous proof, and use the decomposition
\[
\hat\psi^a - \psi^{*a}
=
\hat\psi^a - \mathbb{E}_\Pi[\psi^a]
\;= \;\underbrace{\frac1K\sum_{k=1}^K\bigl(\hat\psi_k^a-\psi_k^a\bigr)}_{\Delta_K^{\mathrm{within}}}
+
\underbrace{\frac1K\sum_{k=1}^K\psi_k^a-\E_\Pi[\psi^a]}_{\Delta_K^{\mathrm{between}}}.
\]
The term $\Delta_K^{\mathrm{between}}$ vanishes by a direct application of the Law of Large Numbers. Let $\bP$ be the set of random variables $(P_k,\mu_k,e_k)_{k \in [K]}$. By boundedness of $\Delta_K^{\rm within}$, we find
\begin{align*}
\bbE[(\Delta_K^{\rm within})^2] &= \bbE[\bbE[(\Delta_K^{\rm within})^2\mid \cE,\bn,\bP]\mid \cE] + o(1) \\ 
&= \frac1{K^2} \sum_{k=1}^K \bbE[\bbE[(\hat\psi_k^a-\psi_k^a)^2\mid \cE,\bn,\bP]\mid \cE] + o(1)  \\
&= \frac{1}{K} \bbE\left[\psi_k^a(1-\psi_k^a)\right]~\bbE\left[\bbE\left[\frac{1}{n_k^a}~\middle|~\cE,\bP\right]~\middle|~\cE\right] + o(1)\\
&\leq \frac{1}{K n \eta_K \ve} + o(1). 
\end{align*}
Since $Kn\eta_K \to \infty$ by assumption, $\Delta_K^{\rm within}$ goes to $0$ in $L^2$, hence in probability.

\subsection{Proof of Proposition~\ref{prp:variance_hetero_pairwise}} \label{app:variance_hetero_pairwise}
We use the vector form of the previous decomposition:
\[
\boldsymbol{\hat\psi} - \boldsymbol{\psi}^{*}
=
\boldsymbol{\hat\psi} - \mathbb{E}_\Pi[\boldsymbol{\psi}]
\;= \;\underbrace{\frac1K\sum_{k=1}^K\bigl(\boldsymbol{\hat\psi}_k-\boldsymbol{\psi}_k\bigr)}_{\boldsymbol{\Delta}_K^{\mathrm{within}}}
+
\underbrace{\frac1K\sum_{k=1}^K\boldsymbol{\psi}_k-\E_\Pi[\boldsymbol{\psi}]}_{\boldsymbol{\Delta}_K^{\mathrm{between}}}.
\]
By the central limit theorem, $\sqrt{K} \boldsymbol{\Delta}_K^{\mathrm{between}}$ goes to $\cN(0,\Sigma)$ with 
$$
\Sigma = \bbE[(\boldsymbol{\psi}_k-\bbE_\Pi[\boldsymbol{\psi}])^\top (\boldsymbol{\psi}_k-\bbE_\Pi[\boldsymbol{\psi}])],
$$
while, by the previous proof,
$$
K \bbE[\|\boldsymbol{\Delta}_K^{\mathrm{within}} \|^2] \leq \frac{1}{n\eta_K \ve} + o(1),
$$
which goes to $0$ by assumption. Hence $\sqrt{K}\boldsymbol{\Delta}_K^{\mathrm{within}} \to 0$, which ends the proof.

\subsection{Proof of Proposition~\ref{prp:variance_hetero}} \label{app:variance_hetero}

Let us introduce 
$$
\bar\bp := \frac1K(|\cK^a|)_{a\in \cA} = \frac1K \sum_{k=1}^K \bp_k \quad \text{with} \quad \bp_k = \left(\ind\{e_k^a > 0\}\right)_{a \in \cA}.
$$

By the law of large number, $\bar\bp$ goes to the positive vector $\bp^{*} = (\Pi(e^a>0))_{a\in\cA}$ as $K \to \infty$. Furthermore,  the decomposition of the previous proof rewrites
in this case
$$
\boldsymbol{\hat\psi} - \boldsymbol{\psi}^{*}
=
 \frac{1}{\bar\bp} \otimes \underbrace{\frac1K\sum_{k=1}^K R_k \otimes \bigl(\boldsymbol{\hat\psi}_k-\boldsymbol{\psi}_k\bigr)}_{\boldsymbol{\wt\Delta}_K^{\mathrm{within}}}
+
\frac{1}{\bar\bp} \otimes  \underbrace{\frac1K\sum_{k=1}^K R_k \otimes \left(\boldsymbol{\psi}_k-\E_\Pi[\boldsymbol{\psi}]\right)}_{\boldsymbol{\wt\Delta}_K^{\mathrm{between}}}.
$$
Just like in the previous case, we can easily prove that $\sqrt{K} \wt\Delta_K^{\rm within} \to 0$ in the regime where $n,K$ and $\eta_K n$ goes to $\infty$. Likewise, we know that
$$
\sqrt{K} \wt\Delta_K^{\rm within} \to \cN(0,\Sigma),
$$
where
$$
\Sigma  = \bbE[(R_k \otimes \left(\boldsymbol{\psi}_k-\E_\Pi[\boldsymbol{\psi}]\right))^\top (R_k \otimes \left(\boldsymbol{\psi}_k-\E_\Pi[\boldsymbol{\psi}]\right))],
$$
and where we used Assumption~\ref{ass:nma_mcar} to have that the random variables $R_k \otimes \left(\boldsymbol{\psi}_k-\E_\Pi[\boldsymbol{\psi}]\right)$ are centered. By Slutsky's lemma, we finally get that
$$
\sqrt{K} (\boldsymbol{\hat\psi} - \boldsymbol{\psi}^{*}) \to \cN\left(0, (\bq^{*\top}\bq^*) \otimes \Sigma\right) \quad \text{where}\quad \bq^* = \frac{1}{\bp^*},
$$
ending the proof.

\end{document}